\documentstyle[12pt]{article}

\begin{document}


\begin{centering}
{\leftskip=2in \rightskip=2in
{\large \bf The three perspectives on the quantum-gravity problem  }}\\
{\leftskip=2in \rightskip=2in
{\large \bf and their implications for the fate of Lorentz
symmetry\footnote{Based on invited seminars given at
{\it Perspectives on Quantum Gravity: a tribute to John Stachel}
(Boston, March 6-7, 2003)
and {\it Tenth Marcel Grossmann Meeting on General Relativity}
(Rio de Janeiro, July 20-26, 2003).}}}\\
\bigskip
\bigskip
\medskip
{\bf Giovanni AMELINO-CAMELIA}\\
\bigskip
{\it Dipart.~Fisica Univ.~La Sapienza and Sez.~Roma1 INFN}\\
{\it P.le Moro 2, I-00185 Roma, Italy}

\end{centering}

\vspace{0.7cm}
\begin{center}
{\bf ABSTRACT}
\end{center}

\baselineskip 11pt plus .5pt minus .5pt

{\leftskip=0.6in \rightskip=0.6in
{\footnotesize Each approach
to the quantum-gravity problem originates
from expertise in one or another area of theoretical physics.
The particle-physics perspective encourages one to attempt to
reproduce in quantum gravity
as much as possible of the successes of the Standard Model of
particle physics,
and therefore, as done in String Theory, the core features
of quantum gravity are described in terms of graviton-like exchange
in a background classical spacetime.
From the general-relativity perspective it is natural to renounce
to any reference to a background spacetime,
and to describe spacetime in a way that takes into account the
in-principle limitations of measurements.
The Loop Quantum Gravity approach and the approaches based on
noncommutative geometry originate from this general-relativity
perspective.
The condensed-matter perspective, which has been adopted in a few
recent quantum-gravity proposals, naturally leads to scenarios
in which some familiar properties of spacetime are only emergent,
just like, for example, some emergent collective degrees of freedom
are relevant to the description of certain physical systems only
near a critical point.
Both from the general-relativity perspective and from
the condensed-matter perspective it is natural
to explore the possibility that quantum gravity might have significant
implications for the fate of Lorentz symmetry
in the Planckian regime.
From the particle-physics perspective there is instead no obvious reason
to renounce to exact Lorentz symmetry, although (``spontaneous'')
Lorentz symmetry breaking is of course possible.
A fast-growing phenomenological programme looking for
Planck-scale departures from Lorentz symmetry
can contribute to this ongoing debate.}}

\newpage
\baselineskip 12pt plus .5pt minus .5pt
\pagenumbering{arabic}

\setcounter{footnote}{0}
\renewcommand{\thefootnote}{\alph{footnote}}

\pagestyle{plain}

\section{Preliminaries}
\subsection{Lorentz symmetry and the
three perspectives on the Quantum Gravity problem}
Each quantum-gravity research line
can be connected with one of
three perspectives on the problem: the particle-physics perspective,
the general-relativity perspective and the condensed-matter perspective.

From a particle-physics perspective it is natural to attempt to
reproduce as much as possible the successes of the Standard Model
of particle physics.
One is tempted to see gravity simply as one more gauge interaction.
Among the quantum-gravity open issues the failure of
perturbative renormalization in naive quantum-gravity is perceived
as most significant.
From this particle-physics perspective a
natural solution of the quantum gravity problem would be String-Theory-like:
a quantum gravity whose core features
are essentially described in terms of graviton-like exchange
in a background classical spacetime.

The general-relativity perspective
naturally leads to reject the use of a background spacetime, and
this is widely acknowledged~\cite{carloreview,leereview}.
Although less publicized, there is also growing awareness
of the fact that the development of general relativity relied heavily
on the careful consideration of the in-principle limitations that
measurement procedures can encounter. Think for example of the limitations
that the speed-of-light limit imposes on certain setups for clock synchronization
and of the contexts in which it is
impossible to distinguish between a constant acceleration and the presence
of a gravitational field.
In light of the various arguments (some briefly reviewed later in these notes)
suggesting that, whenever both quantum mechanics and general relativity
are taken into account, there should be an in-principle
limitation to the localization of a spacetime
point (an event),
the general-relativity perspective
invites one to renounce to any direct reference
to a classical spacetime~\cite{dopl1994,ahlu1994,ng1994,gacmpla,garay}.
Indeed this requirement that spacetime be described as fundamentally
nonclassical (``fundamentally quantum''),
that the in-principle measurability limitations be reflected by the adoption
of a corresponding measurability-limited description of spacetime,
is another element of intuition which is guiding quantum-gravity
research from the general-relativity perspective.
This naturally leads us to consider certain types of
discretized spacetimes, as in the Loop
Quantum Gravity approach, or noncommutative spacetimes.
Loop Quantum Gravity is also a background-independent approach
and therefore combines both elements of the general-relativity
perspective\footnote{Although it must be noted
that this is actually achieved, so far,
at the price of some possibly concerning compromises. For example,
as stressed by John Stachel and others, one could be concerned of the fact
that most of the Loop-Quantum-Gravity results are obtained preserving
only 3D (space) diffeomorphisms.}.
Noncommutative spacetimes could
be introduced in a background-independent way, as in preliminary
attempts reported in Ref.~\cite{dopl1994} and follow-up work,
but in most studies, for simplicity, noncommutative spacetimes are
adopted as background
spacetimes~\cite{gacmaj,seibIRUV,susskind,dineIRUV,gacluisa}
(leading to an approach which in a sense can be seen as originating
from a hybrid of the
particle-physics perspective and the general-relativity perspective).

The third possibility is a condensed-matter perspective
(see, {\it e.g.}, the research programs of Refs.~\cite{volovik}
and \cite{laugh})
on the quantum-gravity problem,
in which some of the familiar properties of spacetime are
only emergent.
Condensed-matter theorists are used to describe some of
the degrees of freedom that are measured in the laboratory
as collective excitations within
a theoretical framework whose primary description is given
in terms of much different, and often practically unaccessible,
fundamental degrees of freedom.
Close to a critical point some symmetries arise for
the collective-excitations theory,
which do not carry the significance of fundamental symmetries,
and are in fact lost as soon as the theory is probed somewhat
away from the critical point. Notably,
some familiar systems are known to exhibit special-relativistic invariance
in certain limits, even though, at a more fundamental level,
they are described in terms of a nonrelativistic theory.
For a rather general class of fermionic systems one finds~\cite{volovik}
that at low energies, as a Fermi point is approached,
fermions gradually become chiral Weyl fermions, while
bosonic collective modes of the vacuum transform into gauge fields
and gravity.

Clearly from the (relatively new) condensed-matter perspective
on the quantum-gravity problem it is natural to see the familiar
classical continuous Lorentz symmetry only as an approximate
(emergent) symmetry.
Results obtained over the last few years (which are partly reviewed
later in these notes) allow us to formulate a similar expectation
from the general-relativity perspective.
Loop quantum gravity and other discretized-spacetime quantum-gravity
approaches appear to require some departures,
governed by the Planck scale, from the familiar
(continuous) Lorentz symmetry.
And in the study of noncommutative spacetimes some Planck-scale departures
from Lorentz symmetry might be inevitable, since (at least in a large
majority of noncommutative spacetimes) a Lie algebra
is not even the appropriate language for the description of
the symmetries of a noncommutative spacetime
(one must resort to the richer structure of Hopf algebras).

From the particle-physics perspective there is instead no obvious reason
to renounce to exact Lorentz symmetry.
Minkowski classical spacetime is an admissible background spacetime,
and in classical Minkowski there cannot be any {\it a priori} obstruction
for classical Lorentz symmetry.
Still, a breakup of Lorentz symmetry,
in the sense of spontaneous symmetry breaking,
is of course possible.
This possibility has been studied extensively~\cite{susskind,dougnekr}
over the last few years, particularly in String Theory, which is
the most mature
quantum-gravity approach that emerged from the particle-physics
perspective.

\subsection{What do we know about quantum-gravity?}
The theory debate clearly is a confrontation between
very different perspectives
on the quantum-gravity problem. If we had any robust
information on quantum gravity certainly at least some
of these ideas would have been proven to fail.
But after more than 70 years~\cite{stachelearly}
of work on the ``quantum-gravity problem" there is
still not a single measured number whose interpretation
requires advocating ``quantum gravity".

I have so far mentioned the quantum-gravity problem as if
it was a well-established and familiar concept, but
it is perhaps useful to give here
an intuitive characterization of this problem.
The quantum-gravity problem is
sometimes described as a sort of ``human discomfort", as a problem
pertaining to the achievement of
a more satisfactory philosophical worldview.
For example, as motivation for research in quantum gravity
it is sometimes stated that ``quantum theory" (in an appropriate
generalized sense) has turned out to be relevant for the description of
measurement results in all other branches of fundamental physics,
and we therefore must assume that it will eventually be relevant
also for spacetime/gravity physics.
Analogously (and amounting to the same thing), it is sometimes stated
that it is unsatisfactory to have on one side
our present unified quantum-field-theory description
of electromagnetic, weak and strong forces
and on the other
side gravity which is still described in a very different way.
These ``human discomforts" do not of course define a scientific problem,
but actually there is, as emphasized by some, a well-defined scientific
problem which can be naturally called ``quantum-gravity problem".

The scientific problem that can be reasonably
called ``quantum-gravity problem"
is actually the problem of producing numbers (predictions), in
a logically-consistent way, for situations in which both gravity effects
and particle-physics quantum-field-theory effects cannot be neglected.
For example, although we are presently (and for the foreseeable future)
unable to set up and observe collisions between two electrons
each with energy of, say, $10^{50}eV$, our present theories
provide no obstruction for the analysis of such high-energy collisions,
but are unable to produce a logically consistent number for, say,
the probability that such a collision would result in two muons
with certain energies and momenta.
The problem, as I shall try to point out later
in these notes, resides in the fact that quantum field theory implicitly
assumes that gravity effects can be neglected. When the gravity
effects are so large that (from the field-theory perspective)
space geometry evolves significantly on very short time scales,
field theory cannot be consistently applied\footnote{Here the reader
should keep in mind that general relativity governs self-consistently
the spacetime dynamics in terms of (and together with)
the particle dynamics, but particle dynamics
is only defined asymptotically, in the S-matrix sense, in
quantum field theory. During a collision process the, say,
electrons involved are not following any trajectories.
We can associate to them some (however fuzzy) trajectories
only asymptotically, much before and much after the collision.
If one tries to apply general relativity to the formally-classical
trajectories that appear in the path integral formulation of quantum
mechanics, the problem becomes anyway ill defined (and affected by
unremovable divergences) if the energies of the particles are high
enough to induce significant geometrodynamics.}.
Similarly, field theory
runs into trouble when gravity effects are strong enough to
admit the emergence of spacetime singularities ({\it e.g.} black holes).
We are able to get ``numbers" out of quantum field theory
in contexts in which there is a curved static (or slowly varying)
nonsingular space, but fast-varying and/or singular space geometries are
untreatable.

One might argue that $10^{50}eV$ electrons should be the
least of our concerns, since we are never going to be able to produce
and/or observe them, but first of all in cosmology there are some
numbers we should produce that depend on very early times after
the big bang (where we have reason to believe that particles
with extremely high energies were abundant), and, secondly,
the fact that our theories fail to
produce numbers in some contexts which those same theories describe as
accessible (in principle) makes us concerned in general
about the robustness of these theories.
Since we know that new elements would have to be introduced in our
theories for the description of collisions between $10^{50}eV$
electrons
(or for a justification of an
in-principle exclusion of such collisions from
the list of processes that can occur in Nature),
it is natural then to wonder whether those new elements can affect
also some of the contexts in which our present theories do provide us
an apparently acceptable prediction.
In some cases the issues we encounter in analyzing, say,
collisions among $10^{50}eV$ electrons might bring to the surface some
issues that could also modify more ordinary
(but still untested) predictions produced by our theories.

There is a very natural explanation for our lack of
insight on this quantum-gravity problem.
One of the few (perhaps the only) robust hint we have about
quantum gravity is that the energy scale
at which the particle-physics
quantum-field-theory description starts to appear inadequate
is the Planck scale $E_p \sim 10^{28} eV$.
For particles of those energies
and higher the fact that the
Standard Model of particle physics ignores
gravitational effects is clearly unsatisfactory.
And usually the scale
that sets the break point of an effective low-energy theory
is also the scale that sets the magnitude of the new effects
to be expected going beyond the effective low-energy theory.
It is therefore reasonable to expect\footnote{Like all expectations
not fully confirmed by experiments, also the apparently-robust
expectation that quantum-gravity effects
be governed by the Planck scale
should be challenged, and has been challenged occasionally.
In particular, the so-called ``theories with
large extra dimensions"~\cite{led}
provide a scenario for an effective increase
of the size of the characteristic quantum-gravity length scale
(decrease of the quantum-gravity energy scale).
These scenarios are not necessarily ``natural", but they do justify some
prudence concerning the assumptions being made on
the characteristic scale of quantum-gravity effects.}
that ``quantum-gravity corrections"
to our low-energy predictions would be very small, with their magnitude
set by some power of the ratio between
the Planck length ($L_p \sim 10^{-35}m$,
which is the inverse the Planck scale $E_p \sim 10^{28} eV$)
and the (much bigger) wavelength of the particles involved in the process.
So we have good reasons to suspect that the quantum-gravity effects
would be very small (and actually they must be typically small,
since we have not managed to see them yet).
Contemplating the horrifying
smallness of the Planck length the quantum-gravity
community had reached the conviction
(see, {\it e.g.}, Ref.~\cite{chrisreview})
that experimental hints could never be obtained.
If this was true, if this expectation was really robust,
there could not possibly be a ``quantum gravity" scientific programme.
But, on the basis of results obtained
over the last 4 or 5 years, it is now clear
that these pessimistic expectations were based
on incorrect premises: ``quantum-gravity experiments" are possible.
Of course, there is no guarantee that they will ever lead to any actual
discovery, but it is clearly incorrect to adopt the {\it a priori}
assumption that the search of the tiny Planck-scale effects should
be hopeless.

In order to look for quantum-gravity effects it is of course
useful to have some guidance from theories.
While the long history of quantum-gravity research
has not led to experimental facts, it did produce
theories that can valuable both for providing
some guidance to quantum-gravity experiments and for clarifying
some technical difficulties that are encountered
in any theory that attempts to incorporate (as appropriate limits)
both general relativity and the quantum field theory that describes the
Standard Model of particle physics.
Results obtained in String Theory provide encouragement
for the idea that a theory
combining gravity and the Standard Model of particle physics
could admit a perturbative treatment (perturbative renormalizability),
at least in certain contexts in which it might be appropriate
to make reference to a background spacetime.
Before these String-Theory results it appeared that ``quantum gravity"
should in all cases be treated using (to-be-determined)
nonperturbative techniques,
with obvious associated difficulties.
Another example is provided by some results obtained in Loop Quantum Gravity,
which provide encouragement for the idea that a
truly background-spacetime-independent
quantum theory can be constructed. Before this loop-quantum-gravity studies
it appeared that there would be a more profound conflict between the
background-spacetime independence of general relativity and
the fact that quantum field theory assumes from the start a background
spacetime.

Unfortunately, loop quantum gravity is being constructed (so far,
pending work in progress),
as a fundamentally nonperturbative theory, without access to the tools
of perturbative analysis which are so valuable in our efforts to ``produce
numbers". And equally unfortunate is the fact that for String Theory
there is (so far) no genuinely background-independent formulation.
But perhaps these weaknesses should not generate too much concern.
For the phenomenological aspects of the line of analysis advocated
in these notes these two theories and other popular quantum-gravity
approaches simply play the role of toy models. The objective being
pursued is the one of finding the first experimental fact
(or even the first few experimental facts)
about the quantum-gravity problem. And we need some guidance.
Where should we look? The toy models can provide
inspiration. Even if neither of them ended up
providing the full solution of the quantum-gravity problem,
it is still rather plausible
that they might have managed
to capture some genuine feature of the correct theory.
Experiments should tell us if this is the case.

\subsection{Quantum Gravity Phenomenology}
The most difficult aspect of the search of experimental hints relevant
for the quantum-gravity problem is the smallness of the
effects that one would naturally expect to be induced by a quantum gravity.
A key point for this ``Quantum Gravity Phenomenology"~\cite{polonpap}
is that we actually are familiar with ways to gain sensitivity
to very small effects.
For example, our understanding of brownian motion is based on the fact
that the collective result of a large number of tiny microscopic effects
eventually leads to observably large macroscopic effects.
Similarly, our present very accurate bounds on the possibility
of a difference in the masses of the $K^0$ and ${\bar {K}}^0$
neutral kaons (relevant for studies of CPT symmetry)
are at a level of precision (better than $\Delta m_K < 10^{-18} m_K$)
which can only be achieved thanks to the fact
that some signatures associated with a $K^0$/${\bar {K}}^0$
mass difference are actually amplified by a large ordinary-physics
number present in the relevant physical contexts (the ratio between
the average mass of neutral kaons and the
difference in mass of the short-living and long-living
neutral-kaon weak-interactions
eigenstates: $m_K \sim 10^{15} [m_{K_L} - m_{K_S}] $).

Now that quantum-gravity phenomenology has grown
into a research area involving some twenty
research groups around the world, it is amusing to compare quantum-gravity
reviews and grandunification reviews written in the early 1990s.
The quantum-gravity reviews considered physics characterized by
the scale $10^{28}eV$ and were claiming that experiments could
never set useful constraints, and simultaneously the grandunification
reviews went into detailed explanations of how certain grandunification
pictures were being rules out by data on proton stability.
The prediction of proton decay within certain grandunification theories
(theories providing a unified description of electroweak and strong
particle-physics interactions) is really a small effect, suppressed
by the fourth power of the ratio between the mass of the proton and the
grandunification scale, which is only
three orders of magnitude smaller than the Planck scale
($E_{gut} \sim 10^{25}eV$).
In spite of this horrifying suppression,
of order $[m_{proton}/E_{gut}]^4 \sim 10^{-64}$,
with a simple idea we have managed to acquire a remarkable sensitivity
to the possible new effect: the proton lifetime predicted by
grandunification
theories is of order $10^{39}s$ and ``quite a few" generations
of physicists should invest their entire lifetimes staring at a single
proton before its decay, but by managing to keep under observation
a large number of protons (think for example of a situation
in which $10^{33}$ protons are monitored)
our sensitivity to proton decay is significantly increased.
In that context the number of protons is the (ordinary-physics)
dimensionless quantity that works as ``amplifier" of the new-physics
effect.

We should therefore focus our attention~\cite{polonpap}
on experiments which have something to do with spacetime
structure and that host an ordinary-physics dimensionless quantity
large enough that it could amplify
the extremely small effects we are hoping to discover.
The amplifier can be the number of small effects contributing
to the observed signal (as in brownian motion and in
proton-stability studies)
or some other dimensionless ordinary-physics number
(as in studies of a possible
difference in the masses of the $K^0$ and ${\bar {K}}^0$).

Using these general guidelines, a few quantum-gravity
research lines have matured over these past few years.
Later in these notes
I will focus on studies of the fate of Lorentz
symmetry in quantum spacetime, emphasizing the relevance for
observations of gamma rays in astrophysics~\cite{grbgac,billetal},
the relevance for the analysis of the cosmic-ray
spectrum~\cite{kifu,ita,aus,gactp,jaco},
and the relevance for certain observations
involving particle decays~\cite{gacpion,orfeupion}.

Concerning laser-interferometric tests of Planck-scale
effects I will only comment on the ones~\cite{gaclaem} that are
directly relevant for the study of the fate
of Lorentz symmetry in quantum spacetime.
I will not discuss ``spacetime foam" studies based
on laser-interferometry, on which there is a
growing literature
(see, {\it e.g.}, Refs.~\cite{gacgwi,ahlunature,nggwi}).

Similarly I will not discuss here matter-interferometric limits
on Planck-scale effects and limits
on Planck-scale effects obtained using the mentioned sensitivity
to new physics that one finds naturally in the neutral-kaon system.
Matter interferometers and the neutral-kaon system were among the
first contexts to be considered from the perspective of
Planck-scale effects (see, {\it e.g.},
Refs.~\cite{ehns,huetpesk,kostcpt,emln,floreacpt,peri}),
but in these contexts there is not much discussion of possible
implications of Planck-scale departures from Lorentz symmetry,
and actually the connection with Planck-scale/quantum-gravity
physics remains rather indirect\footnote{This is perhaps the
reason why the early studies reported in
Refs.~\cite{ehns,huetpesk,kostcpt,emln,floreacpt,peri}
did not manage to generate the interest of a significant portion
of the quantum-gravity community.}.
Both the analysis of matter interferometers and of the neutral-kaon
system appear to require a proper understanding of Planck-scale-induced
decoherence~\cite{emln,peri,garayPRL}, something which we are still
unable to perform satisfactorily even in the simplest quantum
spacetimes (such as the simplest noncommutative spacetimes).
One must therefore rely on general parametrizations,
whose connection with quantum-gravity theories is rather
indirect.
Similarly, the analysis of the neutral-kaon system requires
an understanding of the fate of CPT symmetry in quantum spacetime,
and this too is
something which we are unable to do rigorously even in the
simplest quantum spacetimes~\cite{gacmaj,gacbucc}.

Finally, to give a tentative complete list of quantum-gravity-phenomenology
topics which I will not discuss,
I should stress that I intend to focus here on the possibility
of a ``genuinely quantum" spacetime. I will discuss
(at least intuitively) the difference
between a quantum gravity with a classical spacetime (background)
and a quantum gravity with a genuinely quantum spacetime.
Interesting ideas about the interplay between gravity and quantum mechanics
which do not require a genuinely quantum spacetime can be found in
Refs.~\cite{anan1,ahluiqgr,gasperiniEP,dharamEP,dharamCOW2,veneziano,lamer,ledexp}.

\subsection{A key issue: should we adopt a fundamentally quantum
spacetime?}
As I already stressed it is rather obvious that from the particle-physics
perspective one would not expect any departures from
Lorentz symmetry and on the contrary from the condensed-matter
perspective Lorentz symmetry is naturally seen only as an
approximate symmetry.
It is instead less obvious what one should expect for
the fate of Lorentz symmetry in quantum-gravity approaches
based on the general-relativity perspective, and in fact
some key insight (leading to the expectation that departures
from  Lorentz symmetry are usually present) has been gained
only very recently, mostly in the study of loop quantum gravity
and certain noncommutative spacetimes.

A key point that needs to be clarified when approaching the
quantum-gravity problem from the general-relativity perspective
is whether or not one should adopt  a ``genuinely
quantum'' spacetime.
This concept will not be defined rigorously here,
but combining various points
and remarks in these notes the reader should get an intuitive
picture of this concept.
A genuinely quantum spacetime is essentially a spacetime
in which an event (a spacetime point) cannot be sharply localized.
I will use Loop Quantum Gravity (the present understanding of Loop
Quantum Gravity) and certain noncommutative spacetimes
as examples of genuinely quantum spacetimes.
In the case in which one might be able to introduce coordinates
for the event, in a quantum spacetime it must be impossible
to determine (in the sense of measurement) all of the
coordinates of an event.

I shall use the familiar relativistic quantum field theory
as an example in which spacetime is not fundamentally quantum.
The position of a generic particle cannot be sharply
determined in relativistic quantum field theory, but
it is possible to determine sharply the position
of a particle with infinite mass.
This infinite-mass limit gives operative meaning to
the classical spacetime background in which we describe
relativistic quantum field theory.
I shall argue that if it was not for this infinite-mass limit,
if there was any incompatibility between the infinite-mass limit
and the logical structure of relativistic quantum field theory,
it would have been impossible to make reference to a classical
background spacetime. But
there is no incompatibility between the infinite-mass limit
and the logical structure of relativistic quantum field theory,
so we do have a classical spacetime in that context.


I will stress that the infinite-mass limit is evidently troublesome
once gravity is taken into account, and
I will argue, summarizing the evidence that
emerged in several studies, that a theory that truly admits
both general relativity and quantum mechanics as appropriate
limits must renounce to any reference to a classical spacetime.
Such a theory is automatically incompatible with
the possibility of localizing sharply a spacetime point.

This point is relevant for the fate of Lorentz symmetry
in quantum gravity.
I will review results obtained over the last 3 or 4 years
which suggest that, if spacetime is ``quantum'' in the sense
of noncommutativity or discreteness, the familiar (classical/Lie-algebra,
continuous) Lorentz symmetry naturally ends up being only an approximate
symmetry of the relevant ``flat-spacetime limit'' of quantum gravity.

\subsection{Outline}
These notes are composed of various sections and each section
is (nearly) self-contained. Only in rare cases there is a direct
reference in a given section to a previous section,
but through the combination of the points made in the different
sections I am trying to provide the different elements of
a certain view of the quantum-gravity problem.

The next section is an aside
on the hypothesis of a ``genuinely quantum" spacetime.
I will argue that there should be an absolute limit
on the localization of an event in quantum gravity,
and that this fact should invite us to renounce to any
reference to a non-physical classical spacetime.

In Section~3 I comment on how different approaches to
the quantum-gravity problem describe the fate of Lorentz
symmetry in quantum gravity.

Section~4 focuses on the fate of Lorentz symmetry in
discretized spacetimes, a topic on which some insight
can be gained on the basis of some rather
general considerations, even without the guidance of
a specific quantum-gravity theory.

In Section~5 I review some recent proposals for testing
scenarios for Planck-scale departures from ordinary (classical,
continuous) Lorentz symmetry.

Some closing remarks are in Section~6.

\section{Aside on the hypothesis of a genuinely quantum spacetime}
\subsection{Classical spacetime and localization}
The concept of a classical spacetime is appropriate
in physics (operatively meaningful)
when the theory of interest allows
to localize sharply a spacetime point.
This statement is intended in the same sense that a classical
concept of angular momentum is only appropriate when the
angular-momentum vector (all of its components) can be sharply
measured. In 19th century physics angular momentum was a classical
concept. In our modern theories we acknowledge the experimental
fact that there are limitations on the measurability of
the angular-momentum vector (one cannot measure all of its
components simultaneously) and therefore we describe angular momentum
using a nonclassical formalism (the one of noncommuting operators)
which captures this measurability limitations.

The consistency between the measurability
limits established by the formalism and the in-principle
measurability limits that affect measurement procedures
is a key requirement for a physical theory.
This important issue usually takes center stage in the physics literature
only when a major ``scientific
revolution'' challenges our understanding of the physical world.
In the course of such a ``revolution" it is natural to question
the logical consistency of the novel theoretical frameworks which
are being proposed. Once these logical-consistency issues
have been settled,
and substantial experimental support for the new theory
has been obtained, the focus shifts toward computational matters: one
is comfortable with the logical structure of the new theory
and with the fact that the new theory has some relevance for
the description of Nature, and therefore
precise calculations and accurate experiments
become the top priority.
For example,
this natural sequence of steps for the development of new theories
is easily recognized in the development of
the ``relativity revolution''
and of the ``quantum-theory revolution''.

While the limited scope of these notes does not allow me to describe
rigorously the issues that are to be considered in measurability
analysis (and its role in establishing the logical consistency
of a formalism), the interested reader can find a careful discussion
in the literature, especially
the literature reporting the debate (among Einstein, Peierls,
Bohr, Rosenfeld and others) on the measurability of the electromagnetic
fields in quantum electrodynamics (see Ref.~\cite{bohrrose} and references
therein). At first these measurability studies appeared to expose an
in-principle limitation on the measurability of electromagnetic fields,
and this was of serious concern since a measurability limit would have
implied an inadequacy of quantum electrodynamics. Quantum electrodynamics
makes direct reference to the electromagnetic fields and describes
them as sharply measurable (although the sharp measurement of a
quantum field can only be achieved at the cost of loosing all information
on a conjugate field). Eventually, it was clarified by Bohr
and Rosenfeld~\cite{bohrrose} that there is no limitation
on the measurability of the electromagnetic fields, and this opened
the way for the wide adoption of quantum electrodynamics.

There are several other examples of the importance of
these studies of in-principle measurability limits,
and of the necessity that the theoretical
framework reproduces faithfully the measurability limits.
In understanding the replacement of absolute time by a relative time
a key role is played by the analysis of how an absolute maximum velocity
limits the synchronization of certain pairs of clocks.
In combing quantum mechanics with special relativity one
must adopt quantum field theory, rather than a
relativistic version of quantum mechanics itself,
because the position of a particle of finite ({\it i.e.} non-infinite)
mass cannot be sharply determined through a logically
consistent measurement procedure.

In ``quantum gravity'', a theory that admits both quantum field theory and
general relativity as appropriate limits, is it legitimate to
adopt a classical spacetime? or is it instead necessary to adopt
a nonclassical description of spacetime? I will argue that spacetime
is fundamentally nonclassical in quantum gravity. And I will argue
that adopting a classical spacetime for quantum gravity is
problematic just in the same sense that
a naive relativistic formulation of quantum mechanics
is problematic (and needs to be replaced by quantum field theory).
The relativistic formulation of quantum mechanics can appear to make
formal sense up to a certain point, but eventually one discovers
some inconsistencies (negative energy states...) and these inconsistencies
are easily traced back to the fact that quantum mechanics assumes that
the position of a particle can be sharply measured, whereas any
procedure that combines the uncertainty principle and special relativity
cannot possibly provide a sharp measurement of the particle position.
I conjecture that analogously any quantum-gravity theory that assumes
a classical spacetime will eventually turn out to be inconsistent
or incomplete, because of the failure to provide a logically-consistent
description of the measurability limits (obtained by combining the
uncertainty principle with general relativity) on the
localization of a spacetime point.

\subsection{Spacetime in classical mechanics and in
nonrelativistic quantum mechanics}
Of course, spacetime is classical in classical mechanics.
This has a precise operative meaning which I shall not
discuss here, since the intuitive picture of a classical
spacetime will suffice for the purposes of these notes.

Spacetime is also classical in nonrelativistic quantum mechanics.
Quantum mechanics introduces an absolute
limit on the (simultaneous) measurability of pairs
of conjugate observables, but each observable can still
be sharply measured (at the cost of loosing all information on a
conjugate observable).
Notably the space coordinates of a particle are independent
observables (not conjugate to one another) so they can all
be measured sharply at once.
Quantum mechanics indeed makes direct reference to a
classical background spacetime. This classical spacetime
can be endowed with proper operative meaning by imagining
a dense array of pointlike synchronized clocks.
The clocks mark the time variable (really an external
variable in quantum mechanics) and the (sharply-measurable)
position of the clocks give
physical meaning to the space coordinates.

I will argue that
the concept of infinite-mass point particles\footnote{Of course,
in analyses aimed at defining operatively certain physical
entities the concept of infinite-mass particle can only
be introduced in the sense of a limiting procedure. For example,
in referring to a dense array of infinite-mass
synchronized clocks one is really thinking of a limiting
procedure in which heavier and heavier clocks are used.}
plays a crucial role in spacetime measurability analysis in
relativistic quantum field theory. Within nonrelativistic
quantum mechanics the role of infinite-mass point particles
is more subtle, and I am not even completely sure that such
particles are necessary.
Still an infinite-mass limit might be hidden in the
discussion of the  dense array of pointlike synchronized clocks,
providing the reference frame. If the clocks had finite mass
one should worry about uncertainties in the time evolution
of the reference frame, due to the fact that position and
velocity cannot be both sharp if the mass is finite.
So it appears that the classical spacetime background of
nonrelativistic quantum mechanics acquires proper
operative meaning only in the limit of infinite mass
of the particles that provide identity to the spacetime
points. This is of course not troublesome since ordinary
quantum mechanics neglects gravitational effects,
and therefore
its logical consistency can rely on the
idealization of a
physical reference frame constituted of infinitely massive
point particles.

\subsection{Spacetime in relativistic quantum field theory}
As discussed in the previous subection, the uncertainty principle
coexists with Galileo relativity (which describes
the symmetries of nonrelativistic quantum mechanics)
in such a way that
a physically meaningful classical spacetime can be introduced.
The formalism of quantum mechanics assumes and requires
a classical spacetime since it describes, in terms of
the wave function, the probability that a particle be found
at time $t$ in the (sharply-defined) space point $(x,y,z)$.

There is no (special-)relativistic version of quantum mechanics
because the interplay of the uncertainty principle and special
relativity does not allow one to consider the probability that a
particle be found at time $t$ in the (sharply-defined) space
point $(x,y,z)$. At time $t$ the particle can only be localized with an
accuracy set by its Compton wavelength. This can be seen by considering
a localization procedure as something which ultimately must
involve an interaction between a probe and the particle under
study. In order for the localization to achieve $\delta x$
accuracy the probe must carry at least energy $1 / \delta x$ (so
that the probe itself is confined to a region of size $\delta x$),
but if this energy $1 / \delta x$ is higher than the mass of the
particle being studied/measured additional copies of the original
particle could be produced (in a special-relativistic framework)
as a result of the measurement
procedure. This is of course incompatible with the idea of using
the procedure for the measurement of the position of a
given particle. The position of a particle of mass $m$ and small
velocity/momentum cannot be measured with better accuracy
than $\delta x \sim 1 / m$. More generally the position
of a particle of energy $E$ cannot be measured with
better accuracy than $\delta x \sim 1 / E$.

The fact that the position of a particle with finite (non-infinite)
mass cannot be sharply determined imposes that instead of
a relativistic quantum mechanics we resort to
(relativistic) quantum field theory.
It is noteworthy that in quantum field theory
it is legitimate to make reference
to a classical spacetime, and indeed quantum field theory
does make reference to a classical spacetime.
In fact, quantum field theory (which again ignores gravity)
is perfectly compatible with the introduction of infinite-mass
point particles, and these can provide a classical spacetime
(a classical reference frame) exactly in the same sense already
discussed above for nonrelativistic quantum mechanics.
For an infinite-mass particle the combination of the uncertainty
principle and special relativity does not introduce a limit
on position measurement (the compton wavelength is $0$).
The (in-priciple) presence of a reference frame constituted of
a network of infinite-mass particles allows us to refer to
a classical spacetime. In that classical spacetime the positions
of finite-mass particles cannot however be sharply determined.

In summary, both ordinary (nonrelativistic) quantum mechanics and
(relativistic) quantum field theory do refer to a classical spacetime
in a logically consistent way.
At least in quantum field theory (but, in the sense discussed
in the preceding subsection, also in nonrelativistic quantum mechanics)
the classical spacetime can be operatively described in terms of
a limiting procedure in which spacetime points are marked by point
particles of larger and larger mass, with the sharp localization
of a spacetime point accessible as the infinite-mass limit of
this procedure. It is therefore possible to
combine special relativity and the uncertainty
principle while preserving a physically meaningful classical spacetime.

This is no longer possible when gravity is present:
combining general relativity and quantum mechanics one finds that
on the one hand sharp localization would still require an infinite
mass point particle, but on the other hand, since general relativity
imposes to treat mass as gravitational charge,
the infinite-mass limit is evidently incompatible with the localization
measurement procedure.

\subsection{Spacetime in quantum gravity: events marked by collisions
involving massless point particles or closed strings}
The idea of an absolute limit on localization has a very long tradition
in quantum-gravity research~\cite{stachelearly}.
Some representative studies of various realizations of
this idea can be found in
Refs.~\cite{dopl1994,ahlu1994,ng1994,gacmpla,mead,padma}.
The simplest argument is found in Refs.~\cite{mead,padma}.
It can be summarized by viewing again
a localization procedure as something which ultimately must
involve an interaction between a probe and the ``target''
particle under study.
Assuming that the probe is a massless particle
(or a massive particle in the
relativistic regime, with velocity high enough to neglect the mass)
one is led straightforwardly to a limit on localization which
is set by the Planck length $L_p$.

Again a key point is that in order for the localization procedure
to achieve $\delta x$
accuracy the probe must carry at least energy $1 / \delta x$.
However, taking now into account gravity we see that it is
necessary to require $\delta x \ge L_p$.
In fact, a source of localization uncertainty comes from
the uncertainties in the gravitational interaction between
the probe and the target.
It suffices to consider these uncertainties in a small
region, of size $\epsilon$, around the collision,
{\it i.e.}
the stage of the procedure when
the probe and the target are at distances of
order $\epsilon$.
And we consider $\epsilon \sim \delta x$ since
in order for the collision to be localized with accuracy $\delta x$
the probe and the target must eventually come to be
at least as close as $\delta x$.
The gravitational energy stored in the system
during this stage of the collision, which lasts a time
of order $\epsilon$ (since the probe is massless/relativistic),
is of order\footnote{For simplicity,
the analysis of the uncertainties in the gravitational
energy stored in the system composed by the photon
and the target particle is here discussed within
Newtonian gravity.
As shown in Ref.~\cite{mead}, the estimate obtained
using Newtonian gravity turns out to be correct
(using general relativity one obtains the same estimate,
after a somewhat more lengthy analysis).}
 $U \sim L_p^2 M E/\epsilon$,
where $M$ is the mass of the target particle and $E$
is the energy of the probe. As a result of the uncertainty
in the probe's energy, $\delta E \sim 1 / \delta x$,
this gravitational energy is also uncertain by an
amount of order $\delta U \sim L_p^2 M \delta E /\epsilon$.
Consequently when the probe-target distance is of order $\epsilon$
the probe momentum is uncertain by an
amount $\delta p_{\gamma} \sim L_p^2 M \delta E /\epsilon$
and by momentum conservation also the momentum of the
target particle has the same
uncertainty $\delta p_{M} \sim L_p^2 M \delta E /\epsilon$.
There is therefore a time interval of order $\epsilon$
around the time of the collision
in which the velocity of the target particle is uncertain
by an amount $\delta v_{M} \sim L_p^2 \delta E /\epsilon$.
One concludes that the position of the collision cannot be
established with better accuracy
than $\delta x' \sim \delta v_{M} \epsilon \sim L_p^2 \delta E$.
This indeed leads to
the conclusion $\delta x \ge L_p$, since $\delta E \sim 1 / \delta x$.
Concerning the uncertainty in the time of the collision one
easily finds (the probe is relativistic) $\delta t \sim \delta x \ge L_p$.
I also observe, in preparation for a point I shall articulate later
in this section, that  $\delta x \delta t \ge L_p^2$.

There has been some interest in
generalizing this analysis to the case
in which the probe is a closed string rather than a photon.
It is useful to observe that
(even though a dedicated study is still missing)
in light of the related findings reported
in Ref.~\cite{venekonmen}
one expects only one relevant difference between
ordinary (point-like) massless probes and
closed-string probes:
closed-string probes have the property that
their size increases with
their momentum in such a way that localization
is limited to $\delta x \ge L_s$, where $L_s$
is the string length, and (since
the analysis in Ref.~\cite{venekonmen} requires $L_s > L_p$)
this is consistent with the general expectation  $\delta x \ge L_p$.

\subsection{Spacetime in quantum gravity:
events marked by collisions of neutral
nonrelativistic particles}
The localization limit $\delta x \ge L_p$ is
widely accepted within the quantum-gravity community.
I stress that this localization limit
relies on two crucial ingredients: the nature and strength
of the gravitational interactions and the fact that a
massless particle with energy uncertainty $\delta E$
has position uncertainty $1/\delta E$.
I observe that,
while in practice localization procedures always rely
on massless (or anyway relativistic probes),
according to the Bohr-Rosenfeld line of
analysis~\cite{bohrrose} (which is really the definitive work on
the role of measurability analyses in the logical structure of a
physical theory) it is necessary to wonder whether
the probes that turn out to be useful for practical reasons are
the ones conceptually best suited for the task of localization.
Moreover, one should not only search among the probes
we find to be available in Nature: any type of (``gedanken'')
probe that is consistent with the
conceptual/formal structure of the theory should be
considered~\cite{bohrrose}.
It is plausible that the correct quantum-gravity would predict
its constituents, but at present, since we are still uncertain
about the structure of the correct theory, any discussion
of a general localization limitation must be accompanied by a very
general analysis of possible probes.

In order to establish a localization limit of more general validity
one should in particular consider the possibility of using
non-relativistic neutral probes,
where ``neutral'' here indicates that they only
carry the familiar gravitational (mass/energy) charge, and
only interact gravitationally.

For neutral non-relativistic probes the generalization
of the analysis reviewed in the preceding section is rather straightforward.
Let me consider the event of collision
between a probe of velocity $V_P \ll 1$ and mass $M_P$ (mass of
the probe) and a target particle of mass $M_T$.
The source of localization uncertainty due to
the uncertainties in the gravitational interaction between
the probe and the target is also easily analyzed in the case
of non-relativistic neutral probe.
Again it suffices to consider these uncertainties in
the stage of the procedure when the probe and the target
have distances of a certain (arbitrarily chosen but small)
order $\epsilon$.
The gravitational energy stored in the system
during this stage of the collision, which lasts a time
of order $\epsilon /V_P$, is of order $U \sim L_p^2 M_T E_P/\epsilon$,
where $E_P$ is the probe's energy. As a result of the
uncertainty $M_P V_P \delta V_P$ in the probe
energy this gravitational energy is also
uncertain\footnote{Note that one can also estimate $\delta U$
as $\delta U \sim [L_p^2 M_T M_P/(\epsilon+\delta x_0)]
- [L_p^2 M_T M_P/\epsilon]$,
thereby obtaining
the result $\delta V_T \ge L_p^2 M_P \delta x_0/(V_P \epsilon^2)$.
This would allow to conclude
that there is a time interval of order $\epsilon/V_P$
around the time of the collision
in which the velocity of the target is uncertain
by an amount $\delta V_T \sim L_p^2 M_P \delta V_P /\epsilon$
and the velocity of the probe is (see above)
uncertain by at least $\delta V_P \ge M_P^{-1} V_P^{-1}$.
From these uncertainties one would again be able to conclude
that the position of the collision can only be established with
accuracy worse than $L_p$.}
by an
amount of order $\delta U \sim L_p^2 M_T M_P V_P \delta V_P /\epsilon$.
Consequently when the probe-target distance is of order $\epsilon$
the probe momentum is uncertain by an
amount $L_p^2 M_T M_P \delta V_P /\epsilon$
and by momentum conservation also the momentum of the
target has the same uncertainty.
There is therefore a time interval of order $\epsilon /V_P$
around the time of the collision in which the velocity of the target is
uncertain by an amount $\delta V_T \sim L_p^2 M_P \delta V_P /\epsilon$,
leading to a target-position
uncertainty $\delta x_T \sim L_p^2 M_P \delta V_P$.
Correspondingly the probe-position uncertainty is
of order $\delta x_P \sim 1/(M_P \delta V_P) $.
Since the overall localization uncertainty will receive contributions
both from $\delta x_T$ and $\delta x_P$,
$\delta x \ge \delta x_T + \delta x_P$,
one concludes that $\delta x \ge L_p/\sqrt{V_P}$.
Following analogous reasoning
for the uncertainty in the time of the event
one finds $\delta t \ge L_p/\sqrt{V_P^3}$.
Since consistency with the hypothesis of using a nonrelativistic probe
requires $V_P \ll 1$, we can safely conclude that
there is nothing to be gained by
substituting massless probes with neutral nonrelativistic probes.
While using a massless probe one finds $\delta x \ge L_p$, the use
of a neutral nonrelativistic probe leads to an even more severe
limit on localization $\delta x \ge L_p/\sqrt{V_P}$.
I also observe that, while with a massless probe we found
$\delta t \delta x \ge L_p^2$,
using neutral nonrelativistic probes the localizzation procedure
is even more strictly
limited: $\delta t \delta x \ge L_p^2/V_P^2 \gg L_p^2$.

\subsection{Spacetime in quantum gravity:
events marked by D-particle collisions}
As stressed above there is a quantum-gravity intuition
that assumes a minimum localization uncertainty $\delta x \ge L_p$.
I have also stressed that this intuition
is based on an argument which is not fully robust,
since it relies on the assumption that relativistic probes
carrying only gravitational charge (``neutral'')
should achieve the best localization of a spacetime point.
In the preceding subsection I removed one of these assumptions
by considering a nonrelativistic probe (still neutral),
and found that this cannot be used to improve on $\delta x \ge L_p$.
In this subsection I intend to remove the other key assumption:
I want to explore the possibility that probes carrying other charges
(in addition to the gravitational) might lead to a better localization.
My conclusion will be positive: certain types of charged particles
can be used to obtain an improved (although still limited)
localization.
This might have rather profound implications, since
it modifies the traditional quantum-gravity intuition
favouring $\delta x \ge L_p$.

My observation actually relies on well-established results.
In the study of the String Theory
a new type of spacetime probe was encountered,
the so-called D-particles.
I shall not enter into a detailed description of this
new probes, which however the interested reader
can easily find in the literature (see, {\it e.g.},
Refs.~\cite{dpart1,dpart2,dbrscatt1,dbrscatt2}).
One key point is that the underlying supersymmetry of
the theoretical framework imposes that D-particles,
besides carrying gravitation charge, also carry another
charge associated (in
an appropriate sense) to the gravitational charge
through supersymmetry.
The end result is that~\cite{dbrscatt2}
as long as the distance $d$ between two D-particles
is greater than $\sqrt{v} L_s$
(denoting with $v$ the relative velocity and $L_s$ the string length)
the energy stored in
a two-D-particle system can be described as $U \sim - L_s^6 v^4/d^7$,
up to an (here irrelevant) overall numerical factor of order 1.
This energy law replaces the corresponding Newton
energy law that applies to uncharged
particles.
Another key property of D-particles is the relation between
the D-particle mass $M_D$ the string coupling $g_s$ and
the string length: $M_D \sim g_s^{-1} L_s^{-1}$.
For $g_s \ll 1$ D-particles with appropriately small relative velocity
can be basically treated as ordinary very-weakly-interacting
point particles up to distances
as small as $g_s^{1/3} L_s$, without encountering
comparatively large relative-position quantum
uncertainties.

On the basis of these facts in Refs.~\cite{dbrscatt1,dbrscatt2}
it was argued, within an analysis of D-particle scattering,
that the point of collision between two D-particles can be
localized with uncertainty as low as $\delta x \sim g_s^{1/3} L_s$
(but not lower).
Since in the relevant theoretical framework $g_s \ll 1$ and
the (10-dimensional) Planck length is $L_p \sim g_s^{1/4} L_s$,
this result suggests that a certain level of subPlanckian
localization accuracy is achievable.

Indeed proceeding in complete analogy with the discussion presented
in the previous two subsections, and taking into account the mentioned
D-particle properties, I find that the limit
on localization using a D-particle probe (and target)
is $\delta x \ge g_s^{1/3} L_s \sim g_s^{1/12} L_p$.
I do observe however that
such a level of spatial localization can only be achieved at
the cost of a rather poor level of temporal localization.
In fact, I find (again proceeding in complete analogy with
the discussion presented in the previous two subsections)
that the event is temporally localized with
uncertainty $\delta t \ge g_s^{-1/3} L_s \ge g_s^{-7/12} L_p$,
and, since $g_s \ll 1$, this amounts to an uncertainty limit which is
significantly larger than the usually expected Planck-scale limit.

The example of D-particles suggests that the usual quantum-gravity
expectation that $\delta x \ge L_p$ and $\delta t \ge L_p$
should separately hold might be incorrect.
It is not obvious that D-particles
should be included in the analysis.
The fact that they have emerged in
String Theory is of course not sufficient to conclude
that they should be part of the correct quantum gravity.
But, in the spirit of the Bohr-Rosenfeld ideas, D-particles
might have to
be considered even if they turned out not to exist in Nature.
And actually we are free to look for other types of formal descriptions
of probes which might achieve an even better localization.
On the basis of this observation, I argue that
a satisfactory conclusive analysis of the  quantum-gravity limit
on localization is still missing.
The fact that the infinite-mass limit is no
longer available
in quantum gravity implies that some localization limit must hold,
but additional studies are needed in order to
establish the exact form of the limit.

In this respect I want to venture formulating a conjecture:
for all types of probes that can be introduced in a logically consistent
manner in quantum gravity the following measurability limit will
apply $\delta x \delta t \ge L_p^2$.
So I argue that $\delta x \ge L_p$ and $\delta t \ge L_p$
might not have to hold simultaneously, but for the product
of the two uncertainties one should always
find\footnote{An uncertainty limit even more significant
than the  $\delta x \delta t \ge L_p^2$ might emerge
in contexts in which there is quantum-gravity-induced decoherence,
and finds support in various versions of the Salecker-Wigner inspired
quantum-gravity measurability
bounds~\cite{ng1994,gacmpla}.}  $\delta x \delta t \ge L_p^2$.
This is actually verified in all the contexts I considered,
including the D-particle context, where
the relation\footnote{Yoneya (see, {\it e.g.} Ref.~\cite{yoneya})
has proposed arguments in favour
of the general validity in String Theory
of an uncertainty relation  $\delta x \delta t \ge L_s^2$.
The arguments adopted by Yoneya do not appear to be fully in
the spirit of more traditional measurability analyses (and therefore
it would be important to find additional evidence in support
of  $\delta x \delta t \ge L_s^2$), but it is nonetheless noteworthy
that an uncertainty relation of the type $\delta x \delta t \ge L_s^2$,
derived in perturbative String-Theory frameworks with $L_s > L_p$,
would be again consistent with my more general
conjecture $\delta x \delta t \ge L_p^2$.}
 $\delta x \delta t \ge L_s^2 \gg L_p^2$ holds.

With respect to this conjecture one possible concern could
come from the fact that the analyses that inspired it
are (to a large extent) 1+1-dimensional.
Perhaps a relation of
the type  $\delta x \delta y \delta z \delta t \ge L_p^4$
should also be considered.
It is tempting\footnote{John Stachel has often
expressed a similar intuition.} to
think of a localization limit which basically states that
a given event can only be localized within a four-volume
of a certain fixed size ({\it e.g.} $L_p^4$).
This would also ease concerns about the covariance
of the measurability limit (a four-volume is a rotation/boost
invariant).

\section{Spacetime and Lorentz symmetry in some quantum-gravity
approaches}

\subsection{The three possibilities for the fate
of Lorentz symmetry in quantum gravity}
There are three possibilities for the
fate of Lorentz symmetry in quantum gravity: Lorentz symmetry
remains unmodified
(exact ordinary Lorentz symmetry in the flat-spacetime limit),
Lorentz symmetry is broken, Lorentz symmetry is deformed.

It is of course not difficult to characterize the case
in which Lorentz symmetry is preserved.
In the flat-spacetime limit of quantum gravity
the familiar relations between observations done by different inertial
observers should emerge. If
a given length is at rest and has value $L$ for observer $O$,
an observer $O'$ boosted with velocity $V$
(along the direction defined by the interval being measured)
with respect to $O$ must attribute to that
length the value $L' = \sqrt{1-V^2} L$.
For massless particles all observers agree on the
dispersion relation $E = p$.

We are also all familiar with the concept of ``broken Lorentz symmetry"
that is being encountered and discussed in some quantum-gravity research
lines. It is completely analogous to the  familiar situation
 in which the presence
of a background selects a preferred class of inertial observers.
This is reflected, for example,
in the fact that the dispersion relation for light
travelling in water, in certain crystals, and in
other media is modified. Of course, the existence of crystals
is fully compatible with a theoretical framework that is fundamentally
Lorentz invariant, but in presence of the crystal
the Lorentz invariance is manifest only when
different observers take into account the different form taken by
the tensors that characterize the crystal (or other background/medium)
in their respective reference systems.
If the observers only take into account the transformation rules
for the energy-momentum of the particles involved in a process
the results are not the ones predicted by Lorentz symmetry.
In particular, the dispersion relation between energy and momentum
of a particle depends on the background (and therefore takes different
form in different frames since the background tensors take different
form in different frames).

While the case in which
Lorentz symmetry is preserved and the case in which Lorentz
symmetry is broken are familiar, the third possibility recently
explored in the quantum-gravity literature, the case of deformed
Lorentz symmetry introduced in Ref.~\cite{gacdsr},
is rather new and it might be useful to describe it here intuitively.
It is the idea that in quantum gravity it might be appropriate
to introduce a second observer-independent scale,
a large-energy/small-length
scale, possibly related to the Planck scale.
It would amount to another step of the same type
of the one that connects Galilei Relativity and
Einstein's Special Relativity: whereas in Galilei Relativity
the description of rotation/boost transformations
does not involve any invariant/observer-independent scale,
the observer-independent speed-of-light scale ``$c$'' is encoded in
the Lorentz rotation/boost transformations (which can be viewed
as a $c$-deformation of the Galileo rotation/boost transformations),
and similarly in the case of a deformed Lorentz symmetry
of the type introduced in Ref.~\cite{gacdsr} there are two scales
encoded in the rotation/boost transformations between inertial
observers (observers which are
still indistinguishable, there is no preferred observer).
In addition to the familiar observer-independent velocity scale $c$,
there is a second, length (or inverse-momentum),
observer-independent scale $\lambda$.

In order to provide additional intuition for
the concept of deformed Lorentz symmetry let me consider the
particular much-studied case
in which the deformation involves a new
dispersion relation $m^2 = f(E,p;\lambda)$
with $f(E,p;\lambda) \rightarrow E^2 - p^2$ in the
limit $\lambda \rightarrow 0$ (see, {\it e.g.},
Refs.~\cite{gacdsr,dsrnext,leedsr} and, for some related
follow-up work in cosmology, also see Ref.~\cite{dsrcosmo}).
A modified dispersion relation can also emerge (and commonly
emerges) when Lorentz symmetry is broken, but of course the
role of the modified dispersion relation in the formalism
is very different in the two cases: when Lorentz symmetry is
broken the modified dispersion relation reflects properties
of a background/medium and the laws of boost/rotation transformation
between inertial observers are not modified, while when Lorentz
symmetry is deformed the modified dispersion relation reflects
the properties of some new laws of boost/rotation transformation
between inertial observers.
This comparison provides an invitation to consider
again the analogy with the transition from Galilei Relativity
to Special Relativity.
In Galilei Relativity, which does not have any
relativistic-invariant scale,
the dispersion relation is written
as $E=p^2/(2m)$ (whose structure fulfills the requirements
of dimensional analysis without the need for dimensionful
coefficients). As experimental evidence in favour of Maxwell equations
started to grow, the fact that those equations involve a
special velocity scale appeared to require (since it was assumed that
the validity of the Galilei transformations should not be questioned)
the introduction of a preferred class of inertial observers, {\it i.e.}
the ``ether" background.
Special Relativity introduces the first observer-independent scale,
the velocity scale $c$, its dispersion relation
takes the form $E^2 = c^2 p^2 + c^4 m^2$ (in which $c$ plays a crucial
role for what concerns dimensional analysis), and the presence
of $c$ in Maxwell's equations is now understood not as a manifestation
of the existence of a preferred class of inertial observers but  rather
as a manifestation of the necessity to deform the Galilei
transformations (the Lorentz transformations are a dimensionful
deformation of the Galilei transformations).
Analogously in some recent quantum-gravity research there has
been some interest (see later in these notes) in dispersion relations
of the type $c^4 m^2 =E^2 -  c^2 \vec{p}^2 + f(E,\vec{p}^2;E_p)$
and the fact that these dispersion relations involve an absolute energy
scale, $E_p$, has led to the assumption that
a preferred class of inertial observers should be introduced
in the relevant quantum-gravity scenarios.
But, as I stressed in the papers proposing physical theories with
deformed Lorentz symmetry~\cite{gacdsr}, this assumption is not
necessarily correct:
a modified dispersion relation involving two dimensionful scales
might be a manifestation of new laws of transformation between
inertial observers,
rather than a manifestation of Lorentz-symmetry breaking.

This concludes my brief non-technical review of the concepts
of preserved Lorentz symmetry, broken Lorentz symmetry,
and deformed Lorentz symmetry.
Before closing this subsection I must however introduce
some terminology, which might be useful to
readers interested in finding additional reading material
on this subject. Relativistic theories based on deformed
Lorentz transformations, with two relativistic-invariant scales,
are sometimes called ``doubly-special relativity''
theories~\cite{gacdsr,dsrnext,leedsr}.
In describing a framework with broken Lorentz symmetry
the fact that there is still full invariance under transformations
taking into account of both the background and the particles
energy/momentum is sometimes called ``invariance
under {\underline{observer}}
Lorentz transformations"~\cite{kostebreaking}.
And the fact that in a framework with broken Lorentz symmetry
there are departures from Lorentz symmetry, if one does
not take into account the laws of transformation of the
backgrounds (and simply transforms the
energy-momentum of the particles involved in a process),
is sometimes described as a ``lack of invariance
under {\underline{particle}}
Lorentz transformations"~\cite{kostebreaking}.

\subsection{Aside on the hypothesis of the Planck scale
as a relativistic invariant}
In quantum-gravity proposals it is very common that the Planck
scale (or some related scale, like the string length)
acquires a special role.
One then must understand what are the implications
of the presence of the Planck scale
for Lorentz symmetry.
The analysis is often rather difficult, since the formalisms
used in quantum-gravity research have a very rich structure.
As partial guidance for this type of studies one can resort
to some useful analogies.
The four topics on which I want to comment briefly are:
the possibility of a maximum-velocity scale and its implications
for Galilei transformations,
the possibility of a minimum-wavelength scale
and its implications for Lorentz transformations,
the possibility of angular-momentum discretization scale
(in the sense of ordinary nonrelativistic quantum mechanics)
and its implications for rotation transformations,
and the possibility of a maximum-acceleration scale
and its implications for Lorentz transformations.

The Galilei transformations act on velocities, and therefore
the presence of a maximum-velocity scale naturally invites one
to consider departures from Galilei symmetry.
If different inertial observers attribute different values
to the maximum-velocity scale one should find that Galilei
symmetry is broken (ether).
If the maximum-velocity scale takes the same value for
all inertial observers then Galilei transformation must be
replaced by new laws of transformation between inertial observers
(Special Relativity), in which
the maximum-velocity scale should appear explicitly
in the formulas that governs the transformations between
different inertial observers (in fact, $c$ does appear
in the Lorentz transformation laws).

The Lorentz transformations act on wavelengths, and therefore
the presence of a minimum-wavelength scale
would naturally invite one
to consider departures from Lorentz symmetry.
If different inertial observers attribute different values
to the minimum-wavelength scale one should find that Lorentz
symmetry is broken (``quantum-gravity ether").
If the minimum-wavelength scale takes the same value for
all inertial observers then Lorentz transformations must be
replaced by new laws of transformation between inertial observers
(doubly special relativity~\cite{gacdsr}), in which
the minimum-wavelength scale should appear explicitly
in the formulas that governs the transformations between
different inertial observers (in fact, the minimum-wavelength
scale does appear
in the deformed transformation laws, in the
doubly-special-relativity frameworks that do assume a
minimum wavelength~\cite{gacdsr}).

Space-rotation transformations do act on the angular-momentum
vector, but, as discussed more carefully in the later Section~5,
the type of $\hbar$-discretization of angular momentum
introduced in ordinary nonrelativistic quantum mechanics
is not affected by the discretization scale~\cite{simonecarlo,areanew}.
Therefore this angular-momentum discretization scale $\hbar$
does not require departures
from space-rotation symmetry.
In the formulas describing space-rotation transformations within
ordinary nonrelativistic quantum mechanics
the angular-momentum discretization scale $\hbar$
does not appear.

The Lorentz transformations do not act on accelerations,
and therefore
the presence of a maximum-acceleration scale would not
encourage one to consider departures
from Lorentz symmetry~\cite{schuacc}.
There is no reason for a maximum-velocity scale to appear
in the formulas that govern the transformations between
different inertial observers.

Clearly all of these scales can be introduced as invariants
of some relevant symmetry transformations,
but it is also clear that the nature of these invariants
is somewhat different.
I would like to tentatively propose the terminologies ``trivial
invariant" and ``nontrivial invariant".
The maximum-velocity scale is a nontrivial invariant of
Lorentz boost transformations. And similarly a maximum-wavelength
scale would be a nontrivial invariant of suitable
doubly-special-relativity (deformed Lorentz)
boost transformations.
The angular-momentum-discretization scale encountered in
ordinary nonrelativistic quantum mechanics is a trivial invariant of
space-rotation transformations. And similarly a maximum-acceleration
scale would be a trivial invariant of Lorentz boost transformations.

\subsection{Fuzzyness and Lorentz symmetry}
The description of different possibilities for the fate of Lorentz
symmetry in quantum gravity was (intentionally) rather vague
in the previous subsections. To provide a more precise characterization
one should in particular describe how the symmetries are realized
on (ensembles of) measurements. In particular,
a point which appears to be often overlooked in the literature
concerns the fact that symmetries should also govern
the formulas that describe quantum uncertainty relations, and if
quantum gravity introduces new uncertainty relations the rotation/boost
transformations will have to be applied also to the
uncertainty relations, so that the
overall picture is consistent
with the symmetry principles.

I will try to give an intuitive description of
this concept in this subsection, by discussing
a specific context.
Consider a source which emits ``simultaneously''
a large number of photons (massless particles),
and the photons are such that their energies $E \pm \delta E$
are contained in a certain wide range $E_0-\Delta E \le E \le E_0+\Delta E$,
the range $\Delta E$ being much larger than the average
uncertainties $\delta E$. With  ``simultaneously'' here one of course
must mean a level of simultaneity which is at least compatible with
what we already know about ordinary quantum mechanics:
since the time of emission is uncertain, $\delta t \sim 1/ \delta E$,
the simultaneity cannot be better than $1/ \delta E$.
One class of predictions  coming from (deformed, broken or preserved)
Lorentz symmetry concerns the average times of arrival at a detector.
Unmodified Lorentz symmetry imposes that the average time of arrival
is independent of energy, $t^f_{E_1} = t^f_{E_2}$ for
any $E_1$ and $E_2$.
In a deformed-Lorentz-symmetry or broken-Lorentz-symmetry
scenario one could have instead a certain dependence of the average time
of arrival on the energy (also see later Section~5).

Setting aside the analysis of the average arrival times (for which
the implications of a given symmetry scenario are easily seen),
let us consider the uncertainty in the arrival times.
An interesting hypothesis is that quantum gravity might affect
these uncertainties. Perhaps a group of particles emitted ``simultaneously''
with energy $E_1$ and energy uncertainty $\delta E$ would reach
the detector after a journey of time duration $T$ with a time-of-arrival
uncertainty $\delta t^f = F(E_1,T,\delta t^i; L_p)$,
where $F$ is some function that describes possible energy dependence,
time-of-travel dependence and dependence on the initial time-of-emission
uncertainty. The actual form of the function $F$ should somehow reflect
the symmetries of the theory (in particular, the admissible forms
for the function $F$ are different depending on whether there is
exact classical Lorenz symmetry, deformed Lorentz symmetry or
broken Lorentz symmetry).

\subsection{Spacetime and Lorentz symmetry in
String Theory}
String Theory is the most mature quantum-gravity approach coming
from the particle-physics perspective.
As such it of course attempts to
reproduce as much as possible the successes of quantum field theory,
with gravity seen (to a large extent)
simply as one more gauge interaction.
Although the introduction of extended objects (strings, branes, ...)
leads to subtle elements on novelty, in String Theory
the core features of quantum gravity
are described in terms of graviton-like exchange
in a background classical spacetime.

Indeed String Theory does not lead to spacetime quantization,
at least in the sense that its background spacetime
has been so far described as completely classical.
However, this point is not
fully settled: it has been shown that String Theory
eventually leads to the emergence of fundamental limitations on
the localization of a spacetime event, which are not yet formalized in
a fully satisfactory manner~\cite{wittenPT},
and this might be in conflict with the assumption of a classical
background spacetime.
The Bohr-Rosenfeld consistency criteria are
not yet satisfied: one adopts
a background spacetime which can be
classical, but then the theory itself tells us that the
localization of a spacetime point is affected by
a fundamental limitation.
Clearly the
description of spacetime in String Theory is still being developed,
and requires the analysis of some subtle points.
This logical inconsistency probably tells us that the classical
spacetime background cannot be anything else but a formal
tool, void of operative meaning, which should be eventually
replaced by a physically meaningful spacetime picture in
which no classical-spacetime idealization is assumed.

If eventually there will be a formulation of String Theory
in a background spacetime that is truly quantum, it is likely
(on the basis of the observations reported in these notes)
that Lorentz symmetry will then not be an exact symmetry
of the theory.
If instead somehow a classical spacetime background can be meaningfully
adopted, of course then there would be no {\it a priori} reason to
conjecture departures from Lorentz symmetry:
classical Minkowski spacetime would naturally be an acceptable
background, and a theory in the Minkowski background can be easily
formulated in Lorentz-invariant manner.
Still , it is noteworthy that,
even assuming that it makes sense to consider a classical
background spacetime, the fate of Lorentz symmetry in String Theory
is somewhat uncertain: it has been found that under appropriate
conditions (a vacuum expectation value for certain tensor fields)
Lorentz symmetry is broken in the sense I described above.
In these cases String Theory admits description (in the effective-theory sense)
in terms of field theory in a noncommutative spacetime~\cite{dougnekr}
with most of the studies focusing on the possibility that
the emerging noncommutative spacetime is ``canonical''.
Canonical noncommutative spacetimes are discussed
in Subsection 3.4 and I postpone to that subsection a discussion
of some features of the relevant Lorentz-symmetry breaking.

In summary in String Theory (as presently formulated, admitting classical
backgrounds) it is natural to expect that Lorentz symmetry be preserved.
In some cases (when certain suitable background/``external'' fields are
introduced) this fundamentally Lorentz-invariant theory can experience
Lorentz-symmetry breaking. There has been so far no significant interest
or results on deformation of Lorentz symmetry in String Theory
(see, however, Refs.~\cite{maggioreSTRINGdsr}).

\subsection{Spacetime and Lorentz symmetry in
Loop Quantum Gravity}
Loop Quantum Gravity is the most mature approach to the quantum-gravity
problem that originates from the general-relativity perspective.
As in the case of String Theory, it must be stressed that the
understanding of this rich formalism is still in progress.
As presently understood, Loop Quantum Gravity predicts an inherently
discretized spacetime~\cite{discreteareaLGQ},
and this occurs in a rather compelling way: it is not that one introduces
by hand an {\it a priori} discrete background spacetime; it is rather
a case in which a fully background-independent analysis ultimately
leads, by a sort of self-consistency, to the emergence of
spacetime discretization.
There has been much discussion recently, prompted by the
studies~\cite{grbgac,gampul,mexweave},
of the possibility that this discretization
might lead to broken
Lorentz symmetry.
Although there are
cases in which a discretization is compatible
with the presence of continuous classical
symmetries~\cite{snyder,simonecarlo,areanew},
it is of course natural, when adopting a discretized spacetime,
to put Lorentz symmetry under careful scrutiny.
Arguments presented in Refs.~\cite{gampul,mexweave,thiemLS},
support the idea of broken Lorentz
symmetry in Loop Quantum Gravity.

Moreover, very recently Smolin, Starodubtsev and I proposed~\cite{kodadsr}
(also see the follow-up study in Ref.~\cite{jurekkodadsr})
a mechanism such that Loop Quantum Gravity
would be described at the most fundamental level as a theory that in the
flat-spacetime limit admits deformed Lorentz symmetry.
Our argument originates from the role that certain quantum symmetry groups
have in the Loop-Quantum-Gravity description of spacetime with
a cosmological constant, and observing that in the flat-spacetime limit
(the limit of vanishing cosmological constant)
these quantum groups might not contract to a classical Lie algebra,
but rather contract to a quantum Hopf algebra.

In summary in Loop Quantum Gravity the study of the fate of Lorentz
is still at a preliminary stage. All three possibilities are still
being explored: Lorentz-symmetry preserved, broken or deformed.
It is noteworthy however that until 3 or 4 years ago there was
a nearly general consensus that Loop Quantum Gravity would preserve
Lorentz symmetry, whereas presently the intuition of a majority of experts
has shifted toward the possibility that Lorentz symmetry be broken or
deformed\footnote{Actually it is of course conceivable that Lorentz symmetry
be deformed and broken. This would mean that at the fundamental level the
laws of transformation between inertial observers are
described {\it a la} doubly-special relativity~\cite{gacdsr}
(deformed Lorentz symmetry),
and then, for example, some tensor fields
acquire a vacuum expectation value.}.

\subsection{On the fate of Lorentz symmetry in
canonical noncommutative spacetime}
There has been much recent interest in flat noncommutative spacetimes,
as possible quantum versions of Minkowski spacetime.
Most of the work has focused on various parts of the two-tensor
parameter space
\begin{equation}
\left[x_\mu,x_\nu\right] = i \theta_{\mu \nu}
+ i \gamma^\beta_{\mu \nu} x_\beta ~.
\label{all}
\end{equation}
The assumption that the commutators of spacetime coordinates
would depend on the coordinates at most linearly is
usually adopted for simplicity, but it also
captures a very general intuition: assuming that the Planck
scale governs noncommutativity (and therefore noncommutativity should
disappear in the formal $L_p \rightarrow 0$ limit)
and assuming that the commutators do not involve singular, $1/x^n$, terms
one actually cannot write anything more general than
\begin{equation}
\left[x_\mu,x_\nu\right] = i L_p^2 Q_{\mu \nu}
+ i L_p C^\beta_{\mu \nu} x_\beta ~,
\label{alllp}
\end{equation}
where now the tensors, $Q$ and $C$, are dimensionless.

Most authors actually consider two particular limits~\cite{wessLANGUAGE}:
the ``canonical noncommutative
spacetimes'', with $\gamma^\beta_{\mu \nu} =0$,
\begin{equation}
\left[x_\mu,x_\nu\right] = i \theta_{\mu \nu}
\label{canodef}
\end{equation}
and the ``Lie-algebra noncommutative spacetimes",
with $\theta_{\mu \nu} =0$,
\begin{equation}
\left[x_\mu,x_\nu\right] = i \gamma^\beta_{\mu \nu} x_\beta ~.
\label{liedef}
\end{equation}

Let me start by discussing briefly the fate of Lorentz symmetry
in canonical noncommutative spacetimes.
An intuitive characterization
can be obtained by looking at wave exponentials. The Fourier
theory in canonical noncommutative spacetime is based~\cite{wessLANGUAGE}
on simple wave exponentials $e^{i p^\mu x_\mu}$ and from
the $[x_\mu,x_\nu] = i \theta_{\mu \nu}$
noncommutativity relations one finds that
\begin{equation}
e^{i p^\mu x_\mu} e^{i k^\nu x_\nu}
= e^{-\frac{i}{2} p^\mu
\theta_{\mu \nu} k^\nu} e^{i (p+k)^\mu x_\mu} ~,
\label{expprodcano}
\end{equation}
{\it i.e.} the Fourier parameters $p_\mu$ and $k_\mu$ combine just as
usual, with the only new ingredient of the overall phase factor that
depends on $\theta_{\mu \nu}$.
The fact that momenta combine in the usual way reflects the fact that
the transformation rules for energy-momentum from one
(inertial) observer to another are still the familiar, undeformed,
Lorentz transformation rules. However, the product of wave exponentials
depends on $p^\mu \theta_{\mu \nu} k^\nu$, it depends on the ``orientation"
of the energy-momentum vectors $p^\mu$ and $k^\nu$
with respect to the $\theta_{\mu \nu}$ tensor. This is a first indication
that in these canonical noncommutative spacetimes there is Lorentz symmetry
breaking.
The $\theta_{\mu \nu}$ tensor plays the role of a
background that
identifies a preferred class of inertial observers\footnote{Note that
these remarks apply to canonical noncommutative spacetimes
as studied in the most recent (often String-Theory inspired) literature,
in which $\theta_{\mu \nu}$ is indeed simply a tensor (for a given
observer, an antisymmetric matrix of numbers).
I should stress however that the earliest studies of canonical noncommutative
spacetimes (see Ref.~\cite{dopl1994} and follow-up work)
considered a $\theta_{\mu \nu}$ with richer mathematical properties,
notably with nontrivial algebra relations with the spacetime coordinates.
In that earlier, and more ambitious, setup it is not obvious that Lorentz
symmetry is broken: the fate of Lorentz symmetry
depends on the properties attributed to $\theta_{\mu \nu}$.}.
Different particles are affected by the presence of this background
in different ways, leading to the emergence of different
dispersion relations,
as shown by the results~\cite{seibIRUV,susskind,dineIRUV,gacluisa}
of the study of field theories in canonical noncommutative spacetimes.

\subsection{On the fate of Lorentz symmetry
in  $\kappa$-Minkowski noncommutative spacetime}
In canonical noncommutative spacetimes Lorentz symmetry is ``broken''
and there is growing evidence that Lorentz symmetry breaking occurs
for most choices of the tensors $\theta$ and $\gamma$.
It is at this point clear, in light of several recent results,
that the only way to preserve Lorentz symmetry
is the choice $\theta = 0 =\gamma $, {\it i.e.} the case in which
there is no noncommutativity
and one is back to the familiar classical
commutative Minkowski spacetime.
When noncommutativity is present
Lorentz symmetry is usually
broken, but
recent results suggest that for some special choices of the
tensors $\theta$ and $\gamma$
Lorentz symmetry might be deformed, rather than broken.
In particular, this appears to be the case for the Lie-algebra
$\kappa$-Minkowski~\cite{gacmaj,majrue,kpoinap,lukieFT,gacmich,wesskappa}
noncommutative spacetime ($l,m = 1,2,3$)
\begin{equation}
\left[x_m,t\right] = {i \over \kappa} x_m ~,~~~~\left[x_m, x_l\right] = 0 ~.
\label{kmindef}
\end{equation}

 $\kappa$-Minkowski
is a Lie-algebra spacetime that clearly enjoys classical space-rotation
symmetry; moreover, at least in the Hopf-algebra sense discussed
in Ref.~\cite{gacmich}, $\kappa$-Minkowski
is invariant under noncommutative translations.
Since I am focusing here on Lorentz symmetry,
it is particularly noteworthy that in $\kappa$-Minkowski
boost transformations are necessarily modified~\cite{gacmich}.
A first hint of this comes from the necessity of a deformed
law of composition of momenta, encoded
in the so-called coproduct (a standard structure for a Hopf algebra).
One can see this clearly by considering the Fourier tranform.
It turns out~\cite{gacmaj,lukieFT,majoek} that in the
$\kappa$-Minkowski case the correct formulation of the Fourier theory
requires a suitable ordering prescription
for wave exponentials. From
\begin{equation}
 :e^{i k^\mu x_\mu}: \equiv e^{i k^m x_m} e^{i k^0 x_0}
~,
\label{order}
\end{equation}
as a result of $[x_m,t] = i x_m/\kappa$
(and $[x_m, x_l] = 0$),
it follows that
the wave exponentials combine in a nontrivial way:
\begin{equation}
(:e^{i p^\mu x_\mu}:) (:e^{i k^\nu x_\nu}:) =
:e^{i (p \dot{+} k)^\mu x_\mu}:
\quad.
\label{expprodlie}
\end{equation}
The notation ``$\dot{+}$" here introduced reflects the
behaviour of the mentioned ``coproduct"
composition of momenta in $\kappa$-Minkowski
spacetime:
\begin{equation}
p_\mu \dot{+} k_\mu \equiv \delta_{\mu,0}(p_0+k_0) + (1-\delta_{\mu,0})
(p_\mu +e^{\lambda p_0} k_\mu) ~. \label{coprod}
\end{equation}

As argued in Refs.~\cite{gacdsr} the nonlinearity of the law of composition
of momenta might require an absolute (observer-independent) momentum scale,
just like upon introducing a nonlinear law of composition of velocities
one must introduce the absolute observer-independent scale of
velocity $c$. The inverse of the noncommutativity scale $\lambda$
should play the role of this absolute momentum scale.
This invites one to consider the possibility~\cite{gacdsr}
that the transformation laws for energy-momentum
between different observers would have two invariants, $c$ and $\lambda$.

It is not yet fully established
whether $\kappa$-Minkowski can be the basis for physical
theories with deformed Lorentz symmetry, but very recent works
provide encouragement for this idea~\cite{gacdsr,dsrnext}.
In work that preceded Refs.~\cite{gacdsr}, some
examples of Hopf algebras, the so-called $\kappa$-Poincar\'e algebras,
which could describe
deformed {\underline{infinitesimal}} symmetry transformations for
$\kappa$-Minkowski, had been worked out~\cite{kpoinap},
but it was believed (on the basis of a few attempts~\cite{kpoinnogroup})
that these algebra structures would not be compatible with
a genuine symmetry group of finite transformations.
In Refs.~\cite{gacdsr} it was proposed that one should look
for deformed transformation laws that form a genuine group
and it was shown that one example of the  $\kappa$-Poincar\'e Hopf algebras
previously considered in the mathematical literature
did allow for the emergence of
a group of finite transformations of the energy-momentum
of a particle (while the same is not
true for other examples of these Hopf algebras).
That result amounts to proving that the mathematics
of  $\kappa$-Poincar\'e Hopf algebras (and therefore
possibly $\kappa$-Minkowski)
can meaningfully describe the one-particle sector of a physical theory
in a way that involves deformed Lorentz symmetry.
But it is still unclear whether some $\kappa$-Poincar\'e Hopf algebras
can be used to construct a theory which genuinely enjoys
deformed Lorentz symmetry throughout, including multiparticle systems.

The recipe adopted in the $\kappa$-Poincar\'e literature for
the description of two-particle systems relies
on a law of composition of momenta obtained
through the coproduct sum (\ref{coprod}),
and an action of boosts on the composed momenta induced by the action
on each of the momenta entering the composition.
This has been adopted in the $\kappa$-Poincar\'e literature even
very recently~\cite{lukieNEWdsr}, not withstanding the new
deformed-Lorentz-symmetry perspective proposed in Ref.~\cite{gacdsr}.
From a deformed-Lorentz-symmetry perspective this $\kappa$-Poincar\'e
description of two-particle systems is not acceptable; in fact,
for a particle-producing
collision process $a+b \rightarrow c+d$
laws of the type $(p_a \dot{+} p_b)^\mu = (p_c \dot{+} p_d)^\mu$
are inconsistent with the relevant
laws of transformation for the momenta of the four particles.
The condition $(p_a \dot{+} p_b)^\mu = (p_c \dot{+} p_d)^\mu$
can be imposed in a given inertial frame but it will then be violated
in other inertial frames ({\it i.e.}
$(p_a \dot{+} p_b)^\mu - (p_c \dot{+} p_d)^\mu = 0 \rightarrow
(p_a' \dot{+} p_b')^\mu - (p_c' \dot{+} p_d')^\mu \neq 0$).

So, in summary, in $\kappa$-Minkowski spacetime there are definitely
some departures from Lorentz symmetry, and it appears likely that
these departures could be codified within the deformed-Lorentz-symmetry
(doubly-special relativity) scenario
proposed in Ref.~\cite{gacdsr},
but more work is needed to fully establish
the role of deformed rotation/boost transformations.

\section{Aside on Lorentz symmetry in discrete spacetimes}
\subsection{Introduction and summary of this Section}
Now that I have provided a general picture of the fate of Lorentz symmetry
in quantum gravity, I am basically ready to consider possible experimental
tests that could establish which of these different pictures is correct.
This is discussed in the next section. Before that,
I thought it might be appropriate to devote a few pages to
an aside on one of the key topics of debate
in the quantum-gravity literature:
does spacetime discretization automatically
imply a broken Lorentz symmetry?

It is rather natural for quantum-gravity research to consider
Planck-scale discretization, and this is one of the reasons for the
interest in possible departures from Lorentz symmetry. In fact, most
types of
spacetime discretizations would be clearly incompatible
with the presence of an exact continuous (Lorentz) symmetry.
While this is true, the assumption that ordinary Lorentz symmetry
be only an approximate symmetry in discrete-spacetime pictures
is often made too simplistically:
some quantum-gravity papers rely on the assumption
that in any discretized space it would not be possible to realize
continuous symmetries (and from that it is concluded that
spacetime discretization would necessarily be in conflict
with continuous Lorentz symmetry).
But clearly it is not true that by introducing some element
of discretization in a space one must necessarily renounce
to the presence of continuous symmetries.
There is an example in physics with which we are all familiar:
angular momentum is discretized in ordinary (nonrelativistic)
quantum mechanics but the theory is still consistent with
invariance under space rotations.

While most forms of discretization are incompatible with continuous
symmetries, some discretizations are compatible with continuous
symmetries. It is therefore not possible to assume {\it a priori}
that any scenario for spacetime discretization considered
in the quantum-gravity
literature should lead to departures from Lorentz symmetry.
The fate of Lorentz symmetry should be examined carefully in any
specific discretization scenario.

In the remainder of this section I make
some observations
which could be useful for the
analysis of the implications for Lorentz symmetry of popular
ideas about a quantum-gravity discretization of the spectrum
of the observables length, area, 3-volume, and 4-volume.
I start with some comments
on the somewhat analogous situation
involving space-rotation symmetry and angular-momentum
discretization.
I then consider the possible implications for Lorentz symmetry
of discretization of length, area, 3-volume, and 4-volume.
In the closing subsection I argue that these observations
might be relevant for the analysis of the fate of Lorentz
symmetry in the Loop Quantum Gravity approach.

\subsection{Space-rotation symmetry in classical and ordinary
(nonrelativistic) quantum mechanics}
For the purposes of these notes it is sufficient to focus on the
implications of space-rotation symmetry for the angular momentum 3-vector.
When an observer measures one or more components of the angular momentum
of a classical system
some facts can immediately be deduced about how that same angular
momentum appears to a second observer\footnote{These
remarks, which concern how the same physical process is
described by different observers, characterize {\underline{passive}}
space-rotation symmetry transformations.
The {\underline{active}} transformations instead
connect different processes observed by a single observer.
For example,
in a world with space-rotation symmetry a collection of systems
prepared in a way that does not break that symmetry will have
to enjoy, as an esemble, the same properties along any given
direction ({\it e.g.} the average result of measurements of $L_x$
should coincide with the average result of measurements
of $L_y$).
Another example of manifestation of space-rotation symmetry
within the class of processes observed by a single observer
is the fact that the total angular momentum of an isolated
system does not change in time (space-rotation symmetry
imposes a constraint on the physical processes observed
by a single observer by disallowing processes in which the
total angular momentum of an isolated system is not a constant
of time evolution).}, an observer
whose reference axes are rotated
with respect to the ones of the first observer.
Let us call $(x,y,z)$ the axes of the first observer $O$
and $(x',y',z')$ the axes of the second observer $O'$.
If $O$ measures all three components
of the angular momentum, along the $(x,y,z)$ axes,
everything can be said about all of the components
of that angular momentum along the $(x',y',z')$ axes of $O'$.
The triads $(L_x,L_y,L_z)$ and $(L_{x'},L_{y'},L_{z'})$
are of course different but they are related by a simple
rule of transformation (a space-rotation transformation).
Similarly, if $O$ measures the modulus of the angular momentum vector
everything can be said about how that modulus
appears to a second observer:
the value of the modulus is the same
for both observers.
However, if $O$ measures only the $x$ component of the angular momentum
it is still not possible to predict any of the
components $L_{x'},L_{y'},L_{z'}$ that are most meaningful for $O'$.

Space-rotation symmetry transformations are crucial for
the objectivity (``reality") of
the angular-momentum vector.
Each observer characterizes this vector
by three (real, dimensionful) measured numbers. Each of these numbers
is to be seen as the projection of the objective vector  $\vec{L}$ along
one of the axes of the observer, and, of course, in turn these axes
must be physically identified by the observer. For example, an observer
may choose as ``$x$ axis" the direction of a certain magnetic field
(another vector), and in that case a crucial role is played by
the fact that both in measurement and in theory one can meaningfully
consider the projection $\vec{L} {\cdot} \vec{B}$.
The observable simply denoted by ``$L_x$" in the formalism
inevitably corresponds physically to an observable obtained from
two objective vectors, the angular-momentum vector $\vec{L}$ and
a second vector such as $\vec{B}$. When $\vec{B}$
is known one can set up a measurement procedure
for $L_x \equiv \vec{L} {\cdot} \vec{B}$.
Knowledge of the three components $L_x,L_y,L_z$
({\it i.e.} of the projections along a triplet of
orthogonal directions $\vec{B}^{(i)}$,$\vec{B}^{(j)}$,$\vec{B}^{(k)}$)
is sufficient for predicting the component along any other
given direction.
But the
knowledge of a single component, $L_x$, is not sufficient
to determine how the angular-momentum vector projects along
some other direction.

This observation is rather crucial in understanding how
space-rotation symmetry, a classical continuous
symmetry, can be maintained as a symmetry of systems in
ordinary nonrelativistic quantum mechanics, in which
angular momentum is ``discretized''.
One might, at first sight, be skeptical that
some rules of mechanics that discretize angular momentum
could enjoy a continuous symmetry, but more careful reasoning
quickly leads to the conclusion that there is no {\it a priori}
contradiction between discretization and a continuous symmetry.
In fact, the type of
discretization of angular momentum
which emerges in ordinary non-relativistic quantum mechanics
is fully consistent with classical space-rotation symmetry.

It will be proper~\cite{areanew} to speak of classical symmetries
of a quantum
theory whenever all the
measurements that the quantum theory allows
are still subject to the rules imposed by the classical symmetry.
Certain measurements that are allowed in classical mechanics
are no longer allowed in quantum mechanics, but on those
measurements that are still allowed at the quantum level
the symmetry criteria can fail or succeed
just as in classical mechanics.
It is actually easy to verify that
the presence of classical space-rotation symmetry is
perfectly compatible with the principles of ordinary nonrelativistic
quantum mechanics.

Just as in classical mechanics,
in quantum mechanics the information that $O$ obtains by
measuring the square-modulus  $L^2$ of the angular momentum
is sufficient to establish how that square-modulus
appears to a second observer:
the value of the modulus is the same
for both observers. It happens to be the case that the values of $L^2$
are constrained by quantum mechanics on a discrete spectrum
(while all real positive values are allowed in classical mechanics),
but this of course does not represent an obstruction
for the action of the continuous symmetry
on invariants, such as $L^2$.

When $O$ measures the $x$ component, $L_x$, of the angular momentum
it is still not possible to predict the value of
any of the components
of that angular momentum along the $(x',y',z')$ axes of $O'$.
This is true at the quantum level just as much as it is
true at the classical
level. The fact that quantum mechanics constrains
the values of the observable $L_x$ on a discrete spectrum is
compatible with continuous space-rotation symmetry, simply because
the symmetry does not make predictions relevant for the
single measurement of $L_x$.

In classical physics space-rotation symmetry
also governs the relation between the triple sharp
measurement $(L_x,L_y,L_z)$ made by $O$ and the corresponding
measurement of $(L_{x'},L_{y'},L_{z'})$ made by  $O'$.
This statement is neither true nor false in quantum mechanics.
In fact, quantum mechanics excludes\footnote{Of course,
only the properties of
generic eigenstates are of interest here.
The fact that one could have an eigenstate with $L_x=L_y=L_z=0$,
in the special case $L^2 = 0$, has no implications for my argument.
Also note that the condition $L_x=L_y=L_z=0$ does not involve
the discretization scale $\hbar$ and is space-rotation
invariant both at the classical and the quantum level
($L_x=L_y=L_z=0$ $\rightarrow$ $L_{x'}=L_{y'}=L_{z'}=0$).}
the possibility of simultaneous classical/sharp measurement
of all components of angular momentum.
This prediction of the classical symmetry
is, in a sense, not verifiable in ordinary quantum mechanics,
but it would be improper to say that it fails.

In summary, in quantum mechanics
there is a specific type of ``discretization of angular momentum''
which only affects measurements of space-rotation invariants and
measurements, such as the measurement of a single component $L_x$
of angular momentum, on which space-rotation symmetry makes no
predictions at all (space-rotation symmetry is not such that
one can predict the value of $L_{x'}$ on the basis
of the measurement of $L_x$).
Since space-rotation symmetry does govern the relation between the triple
measurement $(L_x,L_y,L_z)$ made by $O$ and the corresponding
measurement of $(L_{x'},L_{y'},L_{z'})$ made by  $O'$,
it is crucial (for the compatibility betweeen quantum mechanics and
space-rotation symmetry) that according to
quantum mechanics the measurement of one component, say $L_x$,
introduces (in general) a significant uncertainty concerning $L_y$
and $L_z$. If some theory (clearly very different
from quantum mechanics) allowed the simultaneous sharp measurement
of ${{L}}_x$, ${{L}}_y$, ${{L}}_z$
and predicted discrete spectra for them,
then the classical continuous space-rotation symmetry would
inevitably fail to apply.

These observations clarify the deep connection between discretization
and noncommutativity. In a space-rotation-invariant theory
discretization of the spectrum of  $L_x$ requires noncommutativity
of $L_x$ with $L_y$ and $L_z$.
In traditional textbooks the relation between discretization
and noncommutativity is only stressed at the level of formalism
(as a property of the operators we use to formalize the
properties of the relevant measurements), but here I have considered
a direct relation between discretization of measurement results
and noncommutativity of observables (intended as an obstruction for
the simultaneous sharp measurement) in the context of a theory
which, like ordinary quantum mechanics, is compatible with space-rotation
symmetry.

We only need one last (but very important) test before concluding
that the discretization of $L_x$ in quantum mechanics is
truly compatible with space-rotation symmetry.
This is connected with the fact that, as
I stressed, discretization of $L_x$ requires, in presence
of space-rotation symmetry, that whenever $L_x$ is sharply measured
the other components $L_y$ and $L_z$ are affected by significant
uncertainty. There is a risk here of a logical inconsistency:
one must verify that (at least some of) the procedures
that are suitable for the sharp measurement of $L_x$
are not such that they require sharp information on $L_y$
and $L_z$.
Even this test is successful: one can indeed measure sharply $L_x$
without using any knowledge of $L_y$
and $L_z$.
For example,
the Stern-Gerlach setup\footnote{The Stern-Gerlach setup
realizes
physically the projection
of the vector $\vec{L}$ along the direction
of a magnetic field $\vec{B}$. It provides the measurement of $L_x$
in terms of a primary measurement which is a measurement of a corresponding
coordinate of the point of arrival of the particle on a screen.
The value of the measured coordinate is insensitive (even in classical physics)
on the value of $L_y$
and $L_z$.}
measures sharply $L_x$
without using any knowledge of $L_y$
and $L_z$.
Since also this final logical-consistency test is successful,
I conclude that the type of discretization of
angular momentum which is realized in ordinary nonrelativistic quantum
mechanics is fully compatible with classical continuous space-rotation
symmetry.

I will later make use of the
criteria introduced in this subsection
for some considerations on the
possibility that
some form of discretization of lengths,
areas, 3-volumes and 4-volumes
might be compatible with the presence
of classical continuous Lorentz symmetry.
The analogy is very close, but there are some important differences.
A key observation arises there
at the last stage of analysis, the one that here required us
to verify that
one could actually measure sharply $L_x$
without using any knowledge of $L_y$
and $L_z$.

\subsection{Discretization of lengths and Lorentz symmetry}
I now want to explore the possibility that
discretization of lengths might be compatible
with ordinary (classical and continous) Lorentz symmetry.
In setting up this analysis it is useful to start
by considering a very simple procedure for the
measurement of lengths, and examine this procedure ignoring,
for the moment, all the measurability limits imposed by
quantum mechanics (I therefore examine the measurement
procedure as if it was carried through in classical mechanics).
Let us consider
a ruler with extremities marked $A$ and $C$ on a given
spaceship $O$ (the inertial frame $O$).
In order to measure the length of the ruler, $L = AC $,
one places a mirror at $C$ and sends a light signal from $A$
to $C$, which eventually returns at $A$ after reflection by the
mirror at $C$. The length of the ruler will be obtained from
the measurement of the time needed for the two-way
journey $A \rightarrow C \rightarrow A$.
This time of travel is measured by a ``light-clock of
size $d$'', {\it i.e.}
another light beam is bounced back and forth between point $A$
and a point $B$ located at a distance $d$ from $A$ in the direction
orthogonal to $AC$
(if $A$ and $C$ lie on the $x$ axis, $B$ has the same $x$ coordinate
as $A$).
A clock ``tick'' corresponds to each event of
return of the light-clock beam at $A$.
For the clock to be useful the distance $d=AB$ must be known
very accurately, and in order to measure $L$ accurately
it must be that $d \ll L$.

I am assuming that the 3-point system $A$, $B$, $C$, is in rigid
motion with respect to the observer on the spaceship
(with respect to the origin
of the inertial frame $O$), but it is useful not to assume that the
points $A$, $B$, $C$ are at rest. A possible dependence
of the measurement result on the velocity of the $A$-$B$-$C$
system will in fact play an important role in some
observations reported later on.
Let us therefore introduce a velocity  $V$, which is the
velocity of the $A$-$B$-$C$
system with respect to the observer on the spaceship;
specifically, let us take this common velocity of the
points $A$, $B$, $C$ as a 3-vector of modulus $V$,
directed along the $AC$ direction (along the $x$ axis),
pointing away from $A$ and toward $C$.

The velocity $V$ is already relevant at the level of establishing
the calibration of the light-clock.
Since the $AB$ light-clock is moving
with velocity $V$ with respect to  $O$, the observer $O$ sees
the trajectory of the light-clock light beam
as a ``zig-zag" between the moving points $A$ and $B$. [For example,
when bounced back from $B$ toward $A$ the light beam, according to
observer $O$, goes in an oblique direction, and while the light beam
progresses toward $A$, the point $A$ keeps moving with velocity $V$.]
We conclude that, according to observer $O$, each tick of the
light-clock corresponds to a
time $\tau_V = 2 \, \overline{AC}/\sqrt{c^2 - V^2}
= 2 \, d/\sqrt{c^2 - V^2}$.
The dependence on $V$ of the light-clock tick time is easily understood
as a manifestation of time dilatation: if the light-clock is at rest ($V=0$)
each tick corresponds to a time $\tau_0 = 2 \, d/c$,
while for $0<V<c$ the light-clock tick corresponds
to $2 \, d/\sqrt{c^2 - V^2} > \tau_0$.

The velocity $V$ also enters in the relation between the measured
time (the time for the two-way $A \rightarrow C \rightarrow A$ journey)
and the sought length $L$ of the $AC$ ruler.
Since the $AC$ ruler is moving with velocity $V$,
the two parts of the two-way journey of the probe are
of different length.
For the first part of the journey of the probe the fact that the ruler
is moving causes an increase of the duration of the probe's trip toward
the next extremity of the ruler,
while for the second part of the journey
the distance is effectively shortened
by the motion of the ruler.
The first part of the journey ($A \rightarrow C$)
requires that the probe travel a distance $cL/(c-V)$,
while for the second part ($C \rightarrow A$) the distance
is $cL/(c+V)$.
The relation between the time $T$ (given in the ``number of ticks''
form $N \tau_V$) needed
by the probe for its two-way journey and the length $L$ of
the $AC$ ruler is
\begin{equation}
T = N \tau_V = {L \over c-V} + {L \over c+V} = {2 c L \over c^2 - V^2}
~,
\label{relazo2}
\end{equation}
 {\it i.e.}
\begin{equation}
L = {T  (c^2 - V^2)   \over 2 c} =
{N \tau_V  (c^2 - V^2)   \over 2 c} =
N d \sqrt{1 - {V^2 \over c^2}}
~,
\label{relazo2reverse}
\end{equation}
Here the $V$ dependence is
a manifestation of the
FitsGerald-Lorentz length contraction. If the ruler is at rest ($V=0$)
its length $L$ is given by the number of ticks of the clock multiplied by
the size of the light-clock, $L=Nd$,
while for $0<V<c$ one finds a contracted length of the
ruler $N d \sqrt{1 - V^2/c^2} < N d$.

Let us now consider a second spaceship/observer $O'$
moving with velocity $V_0$ with respect to O.
Of course the measurement procedure just described
from the $O$ perspective can be simultaneously witnessed
by $O'$.
However, $O'$ must attribute a different calibration
to the light-clock, since for $O'$ the
velocity of the $A$-$B$-$C$
system is $V' = (V+V_0)/(1+V V_0/c^2)$.
It is easy to verify that for $O'$ each tick of the
light-clock amounts to $2 \, d/\sqrt{c^2 - V'^2}$.
And according to $O'$ the result of the measurement procedure,
the fact that the two-way journey of the probe takes $N$ ticks
of the light-clock, leads to the conclusion that
the ruler has length $L' = N d \sqrt{1 - V'^2/c^2}$.

In summary, a $x$-axis boost corresponding to relative $OO'$ velocity $V_0$
is such that a ruler with $x$-axis velocity $V$ and length $L$ for $O$
is, for $O'$, a ruler with velocity $V' = (V+V_0)/(1+V V_0/c^2)$
and length $L' = L \sqrt{(c^2 - V'^2)/(c^2-V^2)}$.
The way in which a Lorentz boost transforms the length $L$ into the
length $L'$ depends on the velocity of the ruler.
It is also important to notice that the measurement procedure
for $L$, at least as here setup in terms of a primary time measurement,
can only be successful if the velocity of the ruler is known
(from Eq.(\ref{relazo2reverse}) we see
that $L = f(T;V) = T  (c^2 - V^2)/(2 c)$).

Let us first focus on the fact that
the way in which a Lorentz boost transforms the length $L$ into the
length $L'$ depends on the velocity of the ruler.
This can be viewed in analogy with the fact that a space-rotation
around the $z$ axis transforms the $x$-axis component of angular
momentum, $L_x$, into the $x'$-axis component, $L_{x'}$, in a way
that depends on the value of the $y$-axis
component $L_y$: $L_{x'} = \cos (\alpha) L_x + \sin (\alpha) L_y$.
As emphasized in the previous subsection, discretization of
the $L_x$ ($L_{x'}$) spectrum can be compatible with invariance
under arbitrary continuous $\alpha$-angle rotation only
if the instances in which $L_x$ does take a sharp (discrete) value
are such that $L_y$ is affected by an irreducible uncertainty.
This is the case in ordinary quantum mechanics,
where the $L_x$ and $L_y$ observables do not commute.
Analogously, in a quantum-gravity theory
in which the spectrum of lengths
is discrete
compatibility with Lorentz symmetry requires~\cite{areanew} that
the length observable does not commute with the velocity observable,
 {\it i.e.} it requires
that  the instances in which the length of the ruler $L$ does take
a sharp (discrete) value
are such that the velocity $V$ of the ruler is affected
by an irreducible uncertainty.
This follows from the fact that
a Lorentz boost by a velocity $V_0$
leads to a transformation
$\{V,L\} \rightarrow \{V',L'\} \equiv \{(V+V_0)/(1+V V_0/c^2),
L \sqrt{(c^2 - V'^2)/(c^2-V^2)} \}$.
The fact that
a Lorentz boost transforms the length $L$ into the
length $L'$ in a way that is continuous and depends
on the velocity of the ruler
leads to a necessary condition, $[L,V] \neq 0$, for length
discretization to be compatible with ordinary Lorentz
symmetry.

However, even in theories in which $[L,V] \neq 0$ it might
still not be possible
to achieve a logically-consistent scheme for the compatibility
of length discretization with Lorentz symmetry.
An obstruction is suggested from the observation that,
as emphasized above, a sharp measurement of
the length of the ruler $L$ requires that
the velocity of the ruler is known exactly.
In ordinary quantum mechanics $L_x$ discretization is compatible
with space-rotation symmetry because sharp measurements of $L_x$
introduce a large uncertainty in the measurement of $L_y$.
Since information on $L_y$ is not needed for the completion
of $L_x$ measurement procedures (as in the
Stern-Gerlach example) it is perfectly logical to contemplate
contexts in which $L_x$ is sharply measured while $L_y$
is affected by an irreducible uncertainty.
Now we have seen that length discretization could be compatible
with Lorentz symmetry only if a sharp measurement of $L$
introduces a large uncertainty in the measurement of $V$.
But since sharp information on $V$ is needed for the completion
of a sharp $L$ measurement procedure it is puzzling to contemplate
contexts in which $L$ is sharply measured while $V$
is affected by an irreducible uncertainty.

This obstruction represents a serious challenge for the
idea of discrete lengths introduced compatibly with
ordinary Lorentz symmetry.
The obstruction cannot be eliminated within the length-measurement
procedure here adopted: any uncertainty in the velocity of the ruler
would lead to at least some uncertainty in measurement of the length
of the ruler (therefore creating a conflict with hypothesis of
sharp measurement of $L$).
Perhaps one should consider other length-measurement procedures,
but the reader will easily verify that all the commonly considered
length-measurement procedures do require sharp knowledge of the
velocity of the ruler in order to achieve a sharp measurement
of its length.
Moreover, from a conceptual perspective it is puzzling to consider
the possibility that only some very special length-measurement
procedures could achieve sharp results.
In fact, we usually refer to the ``length of the ruler'' as if it
was an intrinsic property of the ruler, verifiable with any of a
large choice of possible equivalent measurement procedures.
If we must consider two classes of length measurements, a ``class A''
that can achieve sharp measurement and a ``class B'' that cannot,
we would be more properly thinking of two different
observables\footnote{An observable is properly
introduced through a specific measurement
procedure. We can attribute several measurement procedures to ``the
same'' observable only when these procedures give
the same results (allowing us to abstract the concept of
length as an intrinsic property of the ruler, rather than
a property of a specific measurement procedure applied to
the ruler).},
one sharply measurable and one with fuzzy properties.

Some of these conceptual issues should be studied in the
future, and the results may affect the perspective here advocated,
but in the meantime it appears necessary to take seriously
the obstruction here encountered.
While waiting for studies of alternative length-measurement
procedures (length-measurement procedures that somehow do not
require knowledge of the speed of the ruler),
it is proper to assume that we do not have a logically-consistent
scenario for introducing length discretization in a way that is
compatible with ordinary Lorentz symmetry.

\subsection{On discretization of area, 3-volume and 4-volume}
The conclusions drawn in the previous subsection
for what concerns length discretization are also applicable
to area discretization and 3-volume discretization.
The action of a Lorentz boost on a given area (3-volume)
depends on the velocity of the surface (3-dimensional object)
whose area (3-volume) is being measured.
And a sharp measurement of an area (3-volume)
requires that the velocity
of the surface (3-dimensional object)
is simultaneously known sharply.
This observation can be easily verified by considering the area
of the triangle defined by 3 mirrors. That area can be measured in terms
of a time-of-travel procedure analogous to the one described in
the previous section.

For areas there has been also much discussion~\cite{carloarea}
of the possibility to measure the area of a metal plate
using an electromagnetic device that keeps a second metal
plate at a small distance $d$ and measures the capacity $C$
of the capacitor formed by the two plates.
The primary measurement would be the capacity, and the sought
area would be evaluated through the relation\footnote{The
presence of $\epsilon_0$ reflects the
simplifying assumption that the measurement be perfomed in
absolute vacuum. This simplification does not affect
the validity of my remarks.} $A = d C / \epsilon_0$.
Also in the case of this area-measurement procedure it is
necessary to assume
that one can measure accurately the velocity\footnote{Although
it is rather marginal with respect to the line of analysis
advocated here, I should stress
that in this area-measurement procedure
based on capacity measurement it is necessary to measure the distance
between the plates:
if $d$ is not known sharply then the relation between $C$ and $A$
becomes fuzzy and the discretization of $A$ may become unobservable.
Assuming that all area measurements
rely on some distance/length measurement one would conclude
that there are inevitable logical inconsistencies in any attempt
to construct a theory in which areas can be sharply measured but
lengths cannot be sharply measured.
It is therefore a high priority for future research on these topics
to establish whether it is possible to devise an area-measurement
procedure that does not rely on any length measurements.} of
the (metallic) surface whose
area is being measured. In fact it is necessary to make sure
that the two surfaces that compose the capacitor are parallel
(constant distance $d$) and that they be centered with respect to one
another. If the second surface (the one that belongs to the measuring
device) is much larger than the surface whose area is being measured
one should be concerned about ``boundary effects" since the
formula $A = d C / \epsilon_0$
actually assume a highly symmetric configuration
(it strictly applies to infinite parallel metallic plates).
If the two surfaces are roughly of the same size any relative velocity
would of course affect the capacity.

The situation for area ad 3-volume therefore appears to
be completely analogous to the one more carefully described here
for what concerns lengths.
Area discretization, intended as the existence of ``states''
in which area is sharp (no uncertainty) and only allowed
to take discrete values,
could be compatible with ordinary Lorentz symmetry only in a theory
in which the sharp measurement of the area requires an
irreducible uncertainty in the measurement of the velocity
of the surface whose area is being
measured. But actually, since in the measurement procedures so far
considered the sharp measurement of areas
requires an equally sharp knowledge of the velocity of the surface,
there appears to be a logical obstruction
for the idea of a discretization of area
that is consistent with ordinary Lorentz symmetry.
Clearly an analogous argument applies to 3-volume discretization.

Somewhat different is the case of 4-volumes. A 4-volume is an
invariant of Lorentz transformations.
There should not be any in-principle obstruction for
introducing a discretization
of 4-volumes in a way that is compatible with ordinary Lorentz symmetry.
I am not really familiar with measurement procedures that
allow the measurement of a 4-volume without resorting to separate
measurements of, say, a 3-volume and a time interval,
but assuming that procedures for the direct measurement
of 4-volumes  can be devised it appears natural to assume
that the outcome of these measurements could be constrained on
a discrete spectrum (when the outcome is sharp)
and that this could be implemented in a way that is fully
compatible with ordinary Lorentz symmetry.

\subsection{Lorentz symmetry and  the type of discretization of
area and 3-volume discussed in the Loop Quantum Gravity literature}
As already mentioned earlier in these notes, the present understanding
of Loop Quantum Gravity involves a discretization of spacetime.
The discretization is such that~\cite{discreteareaLGQ}
the spectra of the area observable and of the 3-volume observable
are discrete.

Since it is not uncommon (although not necessary either, as stressed above)
for a ``discretized space" to be affected by departures from
the symmetries of the continuum limit,
the fact that in the Loop Quantum Gravity literature there has been
much discussion of broken Lorentz symmetry
scenarios~\cite{gampul,mexweave,thiemLS}
and deformed  Lorentz symmetry
scenarios~\cite{kodadsr,jurekkodadsr}
could be interpreted as a manifestation of this discretization.

Indeed, following the line of reasoning advocated in this section
(and in Ref.~\cite{areanew}) one is led to the conclusion that,
while in general discretization of a space does not necessarily
imply loss of all continuous symmetries,
a discretization of area and 3-volume
cannot be introduced compatibly with Lorentz symmetry.
Even if one devises a scenario in which the area
observable does not commute with the surface-velocity
observable, one still should end up finding a logical inconsistency
between discretization of area and Lorentz symmetry.
The inconsistency, as stressed above, originates from the fact that,
when the area observable and the surface-velocity observable
do not commute,
a sharp measurement of the area observable appears to be impossible,
since (at least in the most common
area-measurement procedures, some of which have been considered here)
information on the surface velocity is needed in order to perform
an accurate surface-area measurement.

In giving a physical meaning to the area-discretization results it is
commonly stated in the Loop-Quantum-Gravity literature that the flat surface
of a table (or similar examples) could only take certain discrete values.
In light of the analysis here reported this appears to be inconsistent
with an unmodified Lorentz symmetry.

In light of the present preliminary status of the development of
Loop Quantum Gravity, it is perhaps useful to stress that the
arguments presented here do not completely rule out a compatibility
between area discretization and ordinary Lorentz symmetry.
In order for my argument to be applicable the area discretization
must concerns flat surfaces in a flat spacetime.
It seems to me that at present very little is understood of the physical
interpretation of Loop-Quantum-Gravity area eigenstates.
The familiar Lorentz-transformation formulas for areas (and surface velocity)
assume that the underlying spacetime is flat and the surface is flat.
Perhaps none of the  Loop-Quantum-Gravity area eigenstates
would provide a suitable description of this situation
(although the example adopted in the Loop-Quantum-Gravity literature, making
reference to the flat surface of a table, would suggest it).
None of the results obtained in the  Loop-Quantum-Gravity literature appears
to prove that the area of a flat surface can be measured sharply. Perhaps
only certain specific non-flat surfaces can be measured sharply.
If the area discretization only concerns such non-flat surfaces
then there is no obvious reason for questioning ordinary Lorentz symmetry
on the basis of area discretization.

\section{Experimental searches
of Planck-scale departures from Lorentz symmetry}
Studies of the fate of Lorentz symmetry in quantum gravity provide
an excellent example of quantum-gravity-phenomenology research line.
As discussed in the previous sections, in several (though, of course,
not all) approaches to the quantum-gravity problem one finds some evidence
of departures from the familiar Lorentz symmetry.
Like other effects discussed in the quantum-gravity literature, the
ones associated with departures from Lorentz symmetry are very striking
from a conceptual perspective.
There is a general consensus that some strikingly new effects
should be present in quantum gravity, although
different intuitions for the quantum gravity problem may
lead to favouring one or another of these effects.
As mentioned, it was traditionally believed that even such
strikingly new effects (certainly leading to characteristic signatures)
could not be tested because of their small magnitude, set by the small
ratio between the energy of the particles involved and the Planck energy
scale. Work on quantum-gravity phenomenology has proven that
this old expectation is incorrect.
There is of course no guarantee that ``quantum-gravity experiments"
will ever lead to any actual discovery,
but it is clearly incorrect to adopt the {\it a priori} assumption
that the search of the tiny Planck-scale effects should be hopeless.
This point is very clearly
illustrated in the context of tests of Planck-scale departures
from Lorentz symmetry, on which I focus in this section.

Rather than providing a more general discussion, I intend
to convey my point in a simple way by focusing on the possible
emergence of Planck-scale-modified dispersion relations,
\begin{equation}
E^2= m^2 + \vec{p}^2 + f(\vec{p}^2,E,m;L_p)
~,
\label{eq:disp}
\end{equation}
which are found in the large majority of quantum-gravity-motivated
schemes for deviations from ordinary Lorentz invariance
(see, {\it e.g.},
Refs.~\cite{gacmaj,susskind,grbgac,gampul,thiemLS,kpoinap}).

If the function $f$ is nontrivial\footnote{For example, it would
be pointless to introduce an $f=L_p^2 [E^2 - \vec{p}^2 - m^2]^2$,
since then the dispersion relation (\ref{eq:disp}) would
be equivalent to $E^2= m^2 + \vec{p}^2$.}
and the energy-momentum transformation rules are unmodified (the familiar
Lorentz transformations) then clearly $f$ cannot have the exact
same structure for all inertial observers. In this case
Lorentz symmetry is necessarily ``broken", in the sense clarified earlier
in these notes,
and it is legitimate to assume that, in spite of the deformation
of the dispersion relation,
the rules for energy-momentum conservation would be undeformed.

If instead $f$ does have the exact
same structure for all inertial observers, then necessarily
the laws of transformation between observers must be deformed
(they cannot be the ordinary Lorentz transformation rules).
In this case Lorentz symmetry
is deformed, in the sense of the doubly special relativity~\cite{gacdsr}
discussed earlier in these notes. There is no preferred frame.
The deformation of the laws of transformation between
observers impose that one must also necessarily~\cite{gacdsr} deform
the rules for energy-momentum conservation
(these rules are ``laws of physics"
and must therefore be the same for all inertial observers).

While the case of deformed Lorentz symmetry might exercise a stronger
conceptual appeal (since it does not rely on a preferred class of inertial
observers), for the purposes of this paper it is sufficient to consider
the technically simpler context of broken Lorentz symmetry.
Upon admitting a broken Lorentz symmetry
it becomes legitimate, for example, to adopt
a dispersion relation with leading-order-in-$L_p$ form
\begin{equation}
E^2 \simeq \vec{p}^2 + m^2 - \eta (L_p E)^n \vec{p}^2
~,
\label{eq:displead}
\end{equation}
without modifying the rules for energy-momentum conservation.
In (\ref{eq:displead}) $\eta$ is a phenomenological parameter
of order $1$ (and actually, for simplicity, I will often
implicitly take $\eta = 1$). $n$, the
lowest power of $L_p$ that leads to a nonvanishing
contribution, is model dependent.
In any given noncommutative geometry
one finds a definite value of $n$, and it appears to be equally
easy~\cite{gacdsr,dsrnext,grf03ess}
to construct noncommutative geometries with $n=1$ or with $n=2$.
In Loop Quantum Gravity one might typically
expect~\cite{grf03ess}
to find $n=2$, but certain scenarios~\cite{gampul,leeDispRel}
have been shown\footnote{Note however that the Loop-Quantum-Gravity
scenario of Ref.~\cite{gampul} does not exactly lead to the dispersion
relation (\ref{eq:displead}): for photons ($m=0$) Ref.~\cite{gampul}
describes a polarization-dependent effect (birefringence).}
to lead to $n=1$.

I will use this widely-used scheme for Planck-scale Lorentz-symmetry
breaking, with dispersion relation (\ref{eq:displead}) and unmodified
rules for energy-momentum conservation,
to illustrate how a tiny (Planck-length suppressed) effect
can be observed in certain experimental contexts.
The analysis will also show that
the difference between the case $n \! = \! 1$ and
the case $n \! = \! 2$ is very significant from
a phenomenology perspective. Already with $n \! = \! 1$,
which corresponds to effects that are linearly suppressed by the Planck
length, the correction term in Eq.~(\ref{eq:displead}) is very small:
assuming $\eta \! \simeq \! 1$,
for particles with energy  $E \sim 10^{12} eV$,
some of the highest-energy particles we produce in laboratory,
it represents only a correction of one part in $10^{16}$.
Of course, the case $n \! = \! 2$
pays the even higher price of quadratic suppression by the Planck length
and for $E \! \sim \! 10^{12} eV$ its effects are at the  $10^{-32}$
level.

\subsection{Gamma-ray bursts and Planck-scale-induced in-vacuo dispersion}
A deformation term of the type $L_p^n E^n p^2$ in
the dispersion relation, such as the one in (\ref{eq:displead}),
leads to a small energy dependence of the speed of photons
of order $L_p^n E^n$ (using the relation $v = dE/dp$).
An energy dependence of the speed of photons
of order $L_p^n E^n$ is completely negligible (both for $n=1$ and
for $n =2$) in nearly all physical
contexts, but, at least for $n=1$,
it can be significant~\cite{grbgac,billetal}
in the analysis of short-duration gamma-ray bursts that reach
us from cosmological distances.
For a gamma-ray burst a typical estimate\footnote{Up to 1997 the
distances from the gamma-ray bursters to the Earth were not
established experimentally.
By a suitable analysis of the
gamma-ray-burst ``afterglow"~\cite{grbgac},
it is now possible to establish the distance
from the gamma-ray bursters to the Earth for a significant portion of
all detected bursts. $10^{17} s$ is a rough average of this distance
measurements.} of the time travelled
before reaching our Earth detectors is $T \sim 10^{17} s$.
Microbursts within a burst can have very short duration,
as short as $10^{-4} s$.
We therefore have one of the ``amplifiers" mentioned in Section~1:
the ratio between time travelled by the signal and time structure
in the signal is a (conventional-physics) dimensionless
quantity of order $\sim 10^{17}/10^{-4} = 10^{21}$.
It turns out that this ``amplifier" is sufficient to study
energy dependence of the speed of photons
of order $L_p E$. In fact, some of the photons in these bursts
have energies in the $100 MeV$ range and higher.
For two photons with energy difference of order $\Delta E \sim 100 MeV$
an $L_p \Delta E$ speed difference over a time of travel of $10^{17} s$
leads to a relative time-of-arrival delay of
order $\Delta t \sim \eta T L_p \Delta E \sim 10^{-3} s$.
Such a quantum-gravity-induced time-of-arrival delay
could be revealed~\cite{grbgac,billetal}
upon comparison of the structure of the gamma-ray-burst signal
in different energy channels.

The next generation of gamma-ray telescopes,
such as GLAST~\cite{glast},
will exploit this idea to search for
energy dependence of the speed of photons
of order $L_p E$.

The same analysis leading to a time-of-arrival difference of
order $10^{-3} s$ for the $n=1$ case, leads of course to a much
smaller effect in the case $n=2$ (the case of quadratic suppression
by the Planck length). For $n=2$ the same analysis leads to a
time-of-arrival-difference estimate of order $10^{-23} s$,
which is much beyond the sensitivities achievable with GLAST
and all foreseeable gamma-ray observatories.

Some access to effects characterized by the $n=2$ case
could be gained by exploiting the fact that, according to
current models~\cite{grbNEUTRINOnew},
gamma-ray bursters should also emit a substantial amount of
high-energy neutrinos.
With advanced planned neutrino observatories, such as
ANTARES~\cite{antares}, NEMO~\cite{nemo} and EUSO~\cite{euso},
it should be possible to observe neutrinos with energies
between $10^{14}$ and $10^{19}$ $eV$.
Models of gamma-ray bursters predict in particular a substantial
flux of neutrinos with energies of about $10^{14}$ or $10^{15}$ $eV$.
One could, for example, compare
the times of arrival of these neutrinos emitted by
gamma-ray bursters to the corresponding times of arrival of
low-energy photons. For the case $n=1$ one would predict a
huge time-of-arrival difference ($\Delta t \sim 10^4 s$)
and even for the case $n=2$ the time-of-arrival
difference could be significant ({\it e.g.} $\Delta t \sim 10^{-9} s $)
and possibly within the reach of observatories that could conceivably
be planned for the not-so-distant future.

Current models of gamma-ray bursters also predict some
production of neutrinos
with energies extending to the $10^{19} eV$ level and higher.
For such ultra-energetic neutrinos a comparison of time-of-arrival
differences with respect to soft photons also emitted by the burster
should provide, assuming $n \! = \! 2$,
an even more significant signal, possibly at the
level $\Delta t \! \sim \! 1 s $, which would be
comfortably\footnote{For this strategy relying on
ultra-high-energy neutrinos
the delicate point is clearly not timing, but rather the statistics
(sufficient number of observed neutrinos)
needed to establish a robust experimental result.
Moreover, it appears necessary to understand gamma-ray bursters
well enough to establish whether there are typical
at-the-source time delays.
For example, if the analysis is based
on a time-of-arrival comparison between the first (triggering)
photons detected from the burster and the first neutrinos
detected from the burster it is necessary to establish that
there is no significant at-the-source effect such that
the relevant neutrinos and the relevant photons are emitted
at significantly different times.
The fact that this ``time history"
of the gamma-ray burst must be understood only with precision
of, say, $1 s $ (which is a comfortably large time scale with respect
to the short time scales present in most gamma-ray bursts)
gives us some hope that the needed understanding could be achieved in
the not-so-distant future.} within the realm of timing accuracy of the
relevant observatories.

\subsection{UHE cosmic rays and Planck-scale-modified thresholds}
Let us now consider another significant prediction
that comes from adopting
the dispersion relation (\ref{eq:displead}).
I want to show that also certain types of energy
thresholds for particle-production processes may be sensitive
to the tiny $L_p^n E^n p^2$ modification of
the dispersion relation I am considering.
While in-vacuo dispersion, discussed in the preceding subsection,
only depends on the deformation of the dispersion relation\footnote{The
dispersion relation (\ref{eq:displead}) can also be implemented in
a doubly special relativity (deformed Lorentz symmetry)
scenario~\cite{gacdsr}. The in-vacuo-dispersion analysis discussed
in the preceding subsection applies both to Lorentz-symmetry breaking
and Lorentz-symmetry deformation scenarios adopting (\ref{eq:displead}).
When (\ref{eq:displead}) is adopted in a Lorentz-symmetry deformation
scenario it is necessary~\cite{gacdsr} to consistently modify the laws of
energy-momentum conservation. Therefore the analysis of
Planck-scale-modified thresholds discussed in this subsection,
which assumes unmodified laws of energy-momentum conservation,
does not apply to the scenario in which (\ref{eq:displead})
is adopted in a Lorentz-symmetry deformation scenario.
Planck-scale-modified thresholds are present also in the
case of  Lorentz-symmetry deformation, but there are significant
quantitative differences~\cite{gacdsr}.},
the effects considered in this subsection
also depend on the rules for energy-momentum conservation,
which are not modified
in the Lorentz-symmetry breaking scenario I am considering.

Let us start by considering a collision between
a soft photon of fixed/known energy $\epsilon$
and a high-energy photon of energy $E$.
It is useful to review briefly the usual calculation
of the $E$ threshold for electron-positron pair
production: $\gamma + \gamma \rightarrow e^+ + e^-$.
One can optimize the calculation by starting with the
observation that the photon-photon invariant
evaluated in the lab frame
must be equal to (among other things) the electron-positron
invariant evaluated in the center-of-mass frame:
\begin{equation}
(E+\epsilon)^2 - (P-p)^2 = 4 m_e^2
~.
\label{throne}
\end{equation}
Using the ordinary special-relativistic dispersion relation,
this leads to the ``threshold condition"
\begin{equation}
E \ge E_{th}  = m_e^2/\epsilon
~.
\label{thrtwo}
\end{equation}
Notice that in going from (\ref{throne}) to (\ref{thrtwo}),
using the ordinary dispersion relation, the leading-order
terms of the type $E^2$,$P^2$ have cancelled out, leaving behind the
much smaller (if $\epsilon \ll E$) term of order $E \epsilon$.
This cancellation provides the ``amplifier" needed in quantum-gravity
phenomenology, which in this case can be identified as $E/\epsilon$.
In presence of the Planck-scale departures from Lorentz symmetry
the threshold will be significantly modified if $L_p^n E^n p^2$ is
comparable to (or greater than) $E \epsilon$.
While we normally expect $L_p$-related effects to become significant
when the particles involved have energy $1/L_p$, here for $n=1$ the effect
is already significant when $E \sim (\epsilon/L_p)^{1/2}$,
which can be considerably smaller than $1/L_p$ if $\epsilon$ is small.
Analogously for $n=2$ the effect starts being
significant at $E \sim (\epsilon/L_p^2)^{1/3}$.
In fact, adopting the modified dispersion relation (\ref{eq:displead})
and imposing ordinary (unmodified) energy-momentum conservation
one finds~\cite{gactp} the modified threshold relation
\begin{equation}
E_{th} \epsilon - \eta L_p^n E_{th}^{2+n}
{2^n - 1 \over 2^{2+n}} \simeq m_e^2
~
\label{thrTRE}
\end{equation}
(which again is valid when $E \gg m$ and $E \gg \epsilon$).

Analogous modifications of threshold relations are found for other
processes. In particular, the case of photopion
production, $p + \gamma \rightarrow p + \pi$,
also leads to an analogous result in the case in which the
incoming proton has high energy $E$ while the incoming
photon has energy $\epsilon$ such that $\epsilon \ll E$.
In fact, adopting the modified dispersion relation (\ref{eq:displead})
and imposing ordinary (unmodified) energy-momentum conservation
one finds~\cite{gactp}
the modified threshold relation
\begin{equation}
E_{th} \simeq {(m_p + m_\pi)^2 - m_p^2 \over 4 \epsilon}
- \eta {L_p^n E_{th}^{2+n} \over 4 \epsilon } \left(
{m_p^{1+n} + m_\pi^{1+n} \over (m_p + m_\pi)^{1+n}} -1 \right)
~.
\label{deltaeth}
\end{equation}

This result on the photopion-production threshold is relevant
for the analysis of UHE (ultra-high-energy) cosmic rays.
A characteristic feature of the expected cosmic-ray spectrum,
the so-called ``GZK limit", depends on the evaluation of the
minimum energy required of a cosmic ray in order to produce pions
in collisions with cosmic-microwave-background photons.
According to ordinary Lorentz symmetry this threshold energy
is $E_{th}  \! \simeq  \! 5 {\cdot} 10^{19} eV$
and cosmic rays with energy in
excess of this value should loose the excess
energy through pion production.
Strong interest was generated by the
observation~\cite{kifu,ita,aus,gactp,jaco,alfaro}
that the Planck-scale-modified threshold relation (\ref{deltaeth})
leads, for positive $\eta$, to a
significantly higher estimate of the
threshold energy, an upward shift of the GZK limit.
This would provide a description of the observations of the high-energy
cosmic-ray spectrum reported by AGASA~\cite{agasa},
which can be interpreted as an indication of
a sizeable upward shift of the GZK limit.
Both for the case $n= 1$ and  for the case $n= 2$
the Planck-scale-induced upward shift would be large
enough~\cite{kifu,ita,aus,gactp,jaco,grf03ess,alfaro}
for quantitative agreement with the UHE cosmic-ray observations
reported by AGASA.

There are other plausible theory
explanations for the AGASA ``cosmic-ray puzzle",
and the experimental side must be further explored, since another
cosmic-ray observatory, HIRES, has not confirmed the AGASA results.
The situation will become clearer with planned more powerful
cosmic-ray observatories,
such as the Pierre Auger Observatory, which will soon start taking data.
Still, the possibility that in these cosmic-ray studies
we might be witnessing the first manifestation
of a quantum property of spacetime is of course very exciting;
moreover, whether or not they end up being successful in
describing cosmic-ray observations, these analyses provide another
explicit example of a minute Planck-scale effect that can leave observable
traces in actual data. If Auger ends up establishing that UHE cosmic-ray
data are fully consistent with ordinary Lorentz symmetry this would result
in very significant (Planck-scale) experimental bounds on
quantum-spacetime-induced breakup of Lorentz symmetry not only for
the case $n=1$ but also for the case $n=2$.

In addition to the process $p + \gamma \rightarrow p + \pi$
(and its implications for UHE cosmic rays),
also the process $\gamma + \gamma \rightarrow e^+ + e^-$
has been considered from the point of view of experimental tests.
From the result (\ref{thrTRE}) it follows that, if $n=1$,
for $E \sim 10 TeV$ and $\epsilon \sim 0.01 eV$
the modification of the threshold is significant.
These values of $E$ and $\epsilon$ are relevant for the observation
of multi-$TeV$ photons from certain Blazars~\cite{aus,gactp}.
This high-energy photons travel to us from very far
and they travel in an environment populated by soft photons,
some with energies suitable for acting as targets
for the disappearance of
the hard photon into an electron-positron pair.
Depending on some still-poorly-known properties
(such as the density) of the far-infrared
soft-photon background
the spectrum of multi-$TeV$ photons from certain Blazars
carries information that can be used to test the result (\ref{thrTRE}),
if $\eta$ is positive.
For $n=1$ we are very close~\cite{jaco,newlimit,steckerNEW}
to the sensitivity necessary
to test the Planck-scale effects,
but for $n=2$ the effects are too small for testing in
the foreseeable future.

\subsection{Planck-scale modified decay amplitudes}
The study of certain particle decays provides yet
another possibility to test the idea of broken Lorentz symmetry
at the Planck scale in the way codified by the model I am using in this
section as illustrative example (the phenomenological model
that adopts the modified dispersion relation (\ref{eq:displead}),
with unmodified laws of energy-momentum conservation).
For negative $\eta$ some particles which are stable at low energies
become unstable above a certain energy scale, while for positive $\eta$
some particles
which are unstable at low energies become (nearly) stable
at high energies.

Let me start by considering the case of positive $\eta$, which is
the choice of sign needed in order to push upward the GZK limit
for cosmic rays. I consider
the simple example of the decay of a pion into two photons,
and I focus on the case $n = 1$.
Again it is useful to first review the relevant derivation
within ordinary relativistic kinematics.
One can optimize the calculation by starting with the
observation that the photon-photon invariant in the lab frame
should be equal to the pion
invariant:
\begin{equation}
(E+E')^2 - (\vec{p}+\vec{p}')^2 = m_\pi^2
~.
\label{pionone}
\end{equation}
Using the conventional relativistic
dispersion relation this can be easily turned
into a relation between
the energy $E_\pi$ of the incoming pion, the opening angle $\phi$
between the
outgoing photons, and the energy
$E$ of one of the photons
(the energy $E'$ of the second photon
is of course not independent; it is given by
the difference between the energy
of the pion and the energy of the first photon):
\begin{eqnarray}
\cos(\phi) &\! = \!& {2 E E' - m_\pi^2
\over
2 E E'} ~,
\label{pithreshone}
\end{eqnarray}
where indeed $E' = E_\pi - E$.
Of course,  $\cos(\phi)$ must be $ \leq 1$,
since $\phi$ must be a real physical angle
for all values of $E$.
Note however that typically (unless $E \simeq 0$
or $E \simeq E_\pi$) $m_\pi^2 \ll 2 E E' \sim E_\pi^2/2$
and the equation for $\cos(\phi)$
has the form $\cos(\phi) = (2 E E' - \Delta)/2 E E'$,
with $\Delta = m_\pi^2$.
So the fact that $\cos(\phi) \leq 1$ for all values of $E$
depends only on the fact that $\Delta > 0$, which is automatically
satisfied within ordinary relativistic kinematics through the
prediction $\Delta = m_\pi^2$. A new kinematics predicting
that $\Delta < 0$ for some values of $E$ would have
significant implications.
In order to have a negative $\Delta$
it is sufficient to introduce a relatively small correction,
a correction of order $m_\pi^2$.
This is what happens in the scheme I am considering.
The modified dispersion relation (\ref{eq:displead})
when combined with unmodified energy-momentum conservation,
assuming $n = 1$,
modifies the relation between $\phi$, $E_\pi$ and $E$
according to the formula~\cite{gacpion}
\begin{eqnarray}
\cos(\phi) &\! \simeq \!& {2 E E' - m_\pi^2
+ 3 L_p E_\pi E E'
\over
2 E E' + L_p E_\pi E E'} ~.
\label{pithresh}
\end{eqnarray}
This relation shows that at high energies
the phase space available to the decay
is reduced by the  Planck-scale correction:
for given value of $E_\pi$ certain values of $E$
that would normally be accessible to the decay are no longer
accessible (they would require $cos \theta > 1$).
This effect starts to be noticeable at pion energies of
order $(m_\pi^2/L_p)^{1/3} \sim 10^{15}eV$, but only
very gradually (at first only a small portion of the available
phase space is excluded).

As observed in Refs.~\cite{dedenko,gacpion}
this prediction can be tested through its implications
for the longitudinal development of the air showers produced by
interaction of high-energy cosmic-rays with the atmosphere.
The pion lifetime is in fact a key factor in determining
the longitudinal development of these air showers.
Remarkably, certain
puzzling features have been reported
in analyses~\cite{dedenko}
of the longitudinal development of these air showers,
and a possible explanation could be provided~\cite{gacpion} by the
type of high-energy pion stability that the Planck-scale
effects can induce.
Independently of whether or not this preliminary experimental
encouragement is confirmed by more refined data on pion decay,
it is important for the line of argument here presented
that this scheme for the analysis of pion stability is another
example of a Planck-scale effect that can
become significant in processes involving particles
with energies well below the Planck scale.

A interesting result is found also in the case of negative $\eta$:
some particles which are stable at low energies become
unstable at high energies.
A much studied example is ``photon instability":
the process $\gamma \rightarrow e^+ + e^-$ would be allowed
at high energies if one adopts the
modified dispersion relation (\ref{eq:displead})
and unmodified laws of energy-momentum conservation.
The process $\gamma \rightarrow e^+ + e^-$
can be analyzed in close analogy with the previously
discussed process $\gamma + \gamma \rightarrow e^+ + e^-$.
Assuming $n = 1$, one finds that the
process $\gamma \rightarrow e^+ + e^-$ is allowed when
the photon energy is higher
than $(m_e^2/L_p)^{1/3} \sim 10^{13}eV$.
Observations in astrophysics appear to be in
conflict~\cite{jaco,seth}
with this prediction, and therefore the case
with $n=1$ and negative $\eta$
is ruled out experimentally. More evidence that Planck-scale effects
can be tested
(so much so that some possibilities are being ruled out).

\subsection{Interferometry
and Planck-scale-induced in-vacuo dispersion}
In discussions of possible experimental tests of Planck-scale
(quantum-gravity) modifications of Lorentz symmetry
it is commonly assumed that such tests should rely exclusively
on astrophysics, as in the examples discussed so far in
this section. However, L{\"{a}}mmerzahl and I recently
observed~\cite{gaclaem} that in the foreseeable future (perhaps
a not-so-distant future) Planck-scale
modifications of Lorentz symmetry
could be tested in the controlled laboratory setup of
modern laser interferometers (LIGO/VIRGO-type ground interferometers
or LISA-type space interferometers).
Our observation is based on the idea of operating such an interferometer
with two different frequencies\footnote{Modern interferometers
achieve remarkable
accuracies also thanks to an optimization of all experimental
devices for response to light of a single frequency.
The requirement of operating with light at two different
frequencies is certainly a challenge for the realization
of interferometric setups of
the type proposed in Ref.~\cite{gaclaem}.
This and other practical concerns are not discussed here.
The interested reader can find a preliminary discussion
of these challenges in Ref.~\cite{gaclaem}.},
perhaps obtained from a single laser
beam by use of
a ``frequency doubler'' (see {\it e.g.} \cite{Sauter96}).

I here just want to discuss a rough and simple-minded
estimate of the magnitude of the effect,
within a specific interferometric setup.
The interested reader can find a more realistic analysis,
and descriptions of other interferometric setups in
Ref.~\cite{gaclaem}.
Once again I will adopt as illustrative example of Planck-scale
departures from Lorentz symmetry the one codified in
the modified dispersion relation (\ref{eq:displead}).
I will only consider the case $n=1$, {\it i.e.} departures
from the standard dispersion relation that are only linearly
suppressed by the Planck length.
The even smaller effects that one encounters in the case $n=2$
are clearly beyond the reach of the interferometric studies
considered in Ref.~\cite{gaclaem}.

Let me start by considering an interferometer
with two orthogonal arms,
respectively of length $L$ and $L^\prime$. $L$
and  $L^\prime$ are kept distinct because the signal turns out
to be proportional to $|L - L^\prime|$.
Before entering the interferometer, a monochromatic wave with
frequency $\omega$
goes through a frequency doubler.
Both emerging beams, of frequencies $\omega$ and $2 \omega$,
are then split into a part that goes through the arm
of length $L$ and a part that goes through the arm of length $L^\prime$.
When the beams are finally back (after reflection by mirrors)
at the point where the interference patterns are formed,
one then has access to two interference patterns: one
interference pattern combines two waves of frequency $\omega$
and the other interference pattern is formed combining
analogously two waves of frequencies $2 \omega$.
In an idealized setup (ignoring for example a possible wavelength
dependence in beam-mirror interactions)
the observed intensities
would be governed by
\begin{equation}
I_{\omega} \propto
\frac{1}{2} \left(1
+ \cos\phi_{\omega} \right) \, ,
\quad \phi_{\omega} = k (L^\prime - L) \, ,
\label{phase1}
\end{equation}
\begin{equation}
I_{2 \omega} \propto
\frac{1}{2} \left(1
+ \cos\phi_{2 \omega}\right) \, ,
\quad \phi_{2 \omega} = k^\prime (L^\prime - L) \, ,
\label{phase2}
\end{equation}
where $k^\prime$ is the wavelength associated
with the doubled frequency $2 \omega$.

With the ordinary unmodified
dispersion relation, one has that $k^\prime = 2 k$,
but in the case of our Planck-scale-deformed dispersion relation
\begin{equation}
k^\prime \simeq 2 k + \eta \frac{k^2}{\omega_{p}}
\, ,
\label{kkprime}
\end{equation}
where $\omega_p$ is the Planck frequency ($\omega_p \sim E_p$).

One can then rewrite $\phi_{2 \omega} - \phi_{\omega}$,
from (\ref{phase1}) and
(\ref{phase2}), using again the Planck-scale-deformed dispersion relation
\begin{equation}
\phi_{2 \omega} - \phi_{\omega}
= \omega
\left(1 + \frac{3}{2} \eta \frac{\omega}{\omega_{p}}\right) (L^\prime
- L) \, .
\label{phasediffGAC}
\end{equation}

This phase-difference relation characterizes a key difference
between the standard dispersion relation and the
Planck-scale-deformed dispersion relation in the interferometric
setup here considered.
In the case of the standard classical-spacetime dispersion relation
one expects
a specific type of correlations,
which follow straightforwardly from $k^\prime =2 k$,
between the values of $I_{\omega}$
and $I_{2 \omega}$ for given values of $L^\prime - L$.
For example, clearly one expects that
the intensity $I_{2 \omega}$ of
the wave at frequency $2 \omega$ has a maximum
whenever $L^\prime - L = 2 j \pi/\omega$
(with $j$ any integer number),
and that correspondingly the intensity $I_{\omega}$ of
the wave at frequency $\omega$ has either a maximum
or a minimum.
One therefore predicts,
without any need to establish
the value of $j$, that the
configurations in which there is a maximum of $I_{2 \omega}$
must also be
configurations in which there is
a maximum or a minimum of $I_{\omega}$.
The Planck-scale-modified dispersion relation modifies this
prediction: for example, as codified by Eq.~(\ref{phasediffGAC}),
the modified dispersion relation
predicts that configurations in which there is
a maximum of $I_{2 \omega}$
should be such that $I_{\omega}$
is in the neighborhood but not exactly at
one of its maximum/minimum values.
More precisely, when $L^\prime - L$ is such
that $I_{2 \omega}(L^\prime - L)$ is at a maximum value
the quantum-gravity effect in (\ref{phasediffGAC}) predicts
that $I_{\omega}(L^\prime - L)$ should differ from
a maximum/minimum value of $I_{\omega}$ as if
for being out-of-phase by an
amount $(3 \eta/2) \cdot (L'-L) \omega^2 / \omega_{p}$.

This type of characteristic feature could be looked for by,
for example, taking data at values of $L'-L$ that differ from one another
by small (smaller than $1/\omega$)
amounts in the neighborhood of a value of $L'-L$ that corresponds, say,
to a maximum of $I_{2 \omega}$.
Perhaps, techniques for the active control of mirrors which are already
being used in modern interferometers might be adapted
for this task, and the development of dedicated techniques
does not appear beyond our reach.
In Ref.~\cite{gaclaem} we reported a simple-minded comparison of
the magnitude of the Planck-scale-induced phase difference
to the phase sensitivity of LIGO/VIRGO-type
or LISA-type interferometers.
That comparison provided some encouragement for the idea
of performing these tests in the not-so-distant future,
although several technical challenges are to be
overcome before any attempt of an actual realization of
this type of interferometric setup.

\section{Closing remarks}
There is perhaps something to be learned from looking at
quantum-gravity research in the way I here advocated: exploring
the connection between a given
quantum-gravity approach and the perspective that generated it.
This exercise appears to suggest that what
one finds in a given
quantum-gravity approach might be more directly connected
with the perspective which has been adopted, rather than
with something intrinsic in the quantum-gravity problem.
Many features of quantum-gravity approaches that originate
from the particle-physics perspective, such a String Theory,
simply reflect the intuition that one develops working
with the Standard Model of particle physics.
Analogously quantum-gravity approaches that originate from
the general-relativity perspective or from the condensed-matter
perspective carry a strong trace of intuition developed in
working in those fields.

The expectations concerning the fate of Lorentz symmetry
in quantum gravity appear to be a natural way to discriminate
between the different perspectives. As I stressed here,
it is rather obvious that a particle-physics perspective should
lead to quantum-gravity approaches in which there is no {\it a priori}
reason for departures from ordinary Lorentz symmetry.
And it is equally obvious that instead from the condensed-matter
perspective Lorenz symmetry should only emerge as an
approximate symmetry.
Somewhat more subtle are the indications of the general-relativity
perspective for the fate of Lorentz symmetry, and therefore I
devoted a significant portion of these notes to the general-relativity
perspective. Results obtained
in recent years, some of which I reviewed here, suggest that
also from the general-relativity perspective some departures
from Lorentz symmetry might naturally emerge. If one adopts
a description based fundamentally on noncommutative geometry (as
some aspects of the general-relativity perspective could invite
us to do) departures from Lorentz symmetry are really very natural,
perhaps inevitable. And there is now growing evidence that
also when the general-relativity perspective leads to
discretized-spacetime approaches, as in Loop Quantum Gravity,
departures from Lorentz symmetry are naturally encountered.
I here also presented additional observations, of more general
validity, that favour the presence of departures from Lorentz
symmetry in approaches based on the general-relativity perspective.

Tests of possible Planck-scale departures from Lorentz
symmetry, besides giving us a chance of finding the first
experimental {\underline{facts}} about the quantum-gravity realm,
are therefore also a way to check which one of these three
perspectives should be favoured.
Remarkably, as discussed in Section~6,
there will be several opportunities
in these coming years for experimental searches of
Planck-scale departures from Lorentz symmetry.

In choosing a title for these notes I ended up adopting one
suggesting that there are only three possible perspectives
on the quantum-gravity problem (``{\bf The} three perspectives
on the quantum-gravity problem"),
but it is not unlikely that what we really need
is the discovery of a novel fourth perspective, which
may or may not
be based on a combination of the three
perspectives here considered.


\baselineskip 12pt plus .5pt minus .5pt

\vfil

\end{document}